\documentclass[fleqn,12pt]{article}
\usepackage{amssymb,latexsym}
\usepackage{amsmath, amsthm}
\usepackage[OT2,OT1]{fontenc}
\def\cyr{\fontencoding{OT2}\fontfamily{wncyr}\selectfont}
\usepackage{url}
\usepackage{hyperref}
\input xy
\xyoption{all}
\begin{document}
\title{\bf Conformal Yano-Killing tensor for the Kerr metric
and conserved quantities}
\author{Jacek Jezierski\thanks{Partially supported by grants
   KBN 2 P03B 073 24 and EPSRC: EP/D032091/1. E--mail: \texttt{Jacek.Jezierski@fuw.edu.pl}}
     \ and Maciej {\L}ukasik \\ Department of Mathematical Methods in
Physics, \\ University of Warsaw, ul. Ho\.za 69, 00-681 Warsaw, Poland }
\date{PACS numbers: 11.10.Ef, 04.20.Ha, 11.30.Jj}
\maketitle
{\catcode `\@=11 \global\let\AddToReset=\@addtoreset}
\AddToReset{equation}{section}
\renewcommand{\theequation}{\thesection.\arabic{equation}}

\newtheorem{Definition}{Definition}
\newtheorem{Lemma}{Lemma}
\newtheorem{Theorem}{Theorem}

\newcommand{\gn}[2]{\stackrel{\scriptscriptstyle(#1)}{#2}}
\newcommand{\TBR}{T^{\scriptscriptstyle BR}}
\newcommand{\TEM}{T^{\scriptscriptstyle EM}}
\newcommand{\QBR}{CQ^{\scriptscriptstyle BR}}
\newcommand{\QEM}{CQ^{\scriptscriptstyle EM}}
\newcommand{\QYK}{\Theta}
\newcommand{\dtwo}{\mathbf{\Delta}}
\newcommand{\kolo}[1]{\vphantom{#1}\stackrel{\circ}{#1}\!\vphantom{#1}}
\newcommand{\eq}[1]{(\ref{#1})}
\newcommand{\arxiv}[1]{\url{http://arxiv.org/abs/#1}}
\newcommand{\rd}{\,{\rm d}} 
\newcommand{\tr}{\mathop {\rm tr}\nolimits }
\newcommand{\be}{\begin{equation}}
\newcommand{\ee}{\end{equation}}
\newcommand{\ber}{\begin{eqnarray}}
\newcommand{\eer}{\end{eqnarray}}


\newcounter{mnotecount}[section]
\renewcommand{\themnotecount}{\thesection.\arabic{mnotecount}}
\newcommand{\mnote}[1]
{\protect{\stepcounter{mnotecount}}$^{\mbox{\footnotesize  $
      \bullet$\themnotecount}}$ \marginpar{\raggedright\tiny
    $\!\!\!\!\!\!\,\bullet$\themnotecount: #1} }
\newcommand{\JJ}[1]{\mnote{\textbf{JJ:} #1}}

\begin{abstract}
Properties of (skew-symmetric) conformal Yano--Killing tensors are
reviewed. Explicit forms of three symmetric conformal Killing
tensors in Kerr spacetime are obtained from the Yano--Killing tensor.
The relation between spin-2 fields and solutions to the Maxwell
equations is used in the
construction of a new conserved quantity which is
quadratic in terms of the Weyl tensor. The
formula obtained is similar to the functional obtained from the
 Bel--Robinson tensor and is examined in Kerr spacetime.
 A new interpretation of the conserved quantity obtained is proposed.
\end{abstract}

\section{Introduction}
According to \cite{cykem}
one can define, in terms of spacetime curvature, two kinds of conserved
quantities with the help of conformal Yano--Killing tensors (sometimes called
conformal Killing forms or twistor forms,
see e.g. \cite{Moroianu}, \cite{Semmelmann}, \cite{Stepanow}).
The first kind is linear and the second quadratic with respect to
the Weyl tensor but a basis for both of them is the Maxwell field.
Conserved quantities which are linear with respect to CYK tensor
were investigated many times
(cf. \cite{Abbott-Deser}, 
\cite{Glass-Naber}, 
\cite{Goldberg1}, 
\cite{JJspin2}, \cite{kerrnut}, \cite{cykem}, \cite{Pen1}, 
\cite{Pen-Rin}).
On the other hand, quadratic charges are less known and
have usually been examined in terms of the Bel--Robinson tensor
(see e.g. \cite{Berg}, \cite{Sen}, \cite{Ch-Kl}, \cite{Douglas}).

In \cite{cykem} the following super-tensor was introduced
\be\label{T6} \hspace*{-0.7cm}
 {\cal T}_{\mu\nu\alpha\beta\gamma\delta} = \frac12 \left(
W_{\mu\sigma\alpha\beta} W_\nu{^\sigma}{_{\gamma\delta}} +
W_{\mu\sigma\gamma\delta} W_\nu{^\sigma}{_{\alpha\beta}} +
{W^*}_{\mu\sigma\alpha\beta} {W^*}_\nu{^\sigma}{_{\gamma\delta}}+
{W^*}_{\mu\sigma\gamma\delta} {W^*}_\nu{^\sigma}{_{\alpha\beta}}
\right) \, ,
  \ee
where $W_{\mu\sigma\alpha\beta}$ is a spin-2 field (Weyl tensor) and
by ``$*$'' we denote its dual $W^{*}{}_{\mu\nu\alpha\beta} :=
\frac{1}{2}W_{\mu\nu\rho\sigma}\varepsilon^{\rho\sigma}{}_{\alpha\beta}$
(cf. Section \ref{subspin2} below).
The tensor $\cal T$ has the following algebraic properties:
 \be\label{pt1} {\cal T}_{\mu\nu\alpha\beta\gamma\delta}=
   {\cal T}_{\mu\nu\gamma\delta\alpha\beta}=
   {\cal T}_{(\mu\nu)[\alpha\beta][\gamma\delta]}
    \, , \quad
 {\cal T}_{\mu\nu\alpha\beta\gamma\delta}g^{\mu\nu} = 0 \, , \ee
which are simple consequences of the definition \eq{T6} and
spin-2 field properties (cf. Definition \ref{spin2_df} below).
A contraction of ${\cal T}$ gives the Bel--Robinson tensor:
\[ \TBR_{\mu\nu\alpha\gamma}=
g^{\beta\delta}{\cal T}_{\mu\nu\alpha\beta\gamma\delta}
    \, . \]
Moreover, in \cite{cykem} the following properties
\be\label{pt2}
 \nabla^\mu {\cal T}_{\mu\nu\alpha\beta\gamma\delta} = 0
 \, , \quad {\cal T}_{\mu[\nu\alpha\beta]\gamma\delta}=0 \,  \ee
of the tensor $\cal T$ are shown.
They enable one to formulate the following
\begin{Theorem}\label{th1}
If $P,Q$ are CYK tensors, $X$ is a conformal vector field
and ${\cal T}$ obeys the properties \eq{pt1} and \eq{pt2}
then
\[ \nabla^\mu \left( {\cal T}_{\mu\nu\alpha\beta\gamma\delta}
X^\nu P^{\alpha\beta}Q^{\gamma\delta}\right) =0 \, .\]
\end{Theorem} \noindent
The basis of Theorem \ref{th1} consists in the observation that
the  skew-symmetric tensor $F_{\mu\nu}:=W_{\mu\nu\alpha\beta}Q^{\alpha\beta}$
fulfils Maxwell's equations.

We summarize some relationships by the following diagram:
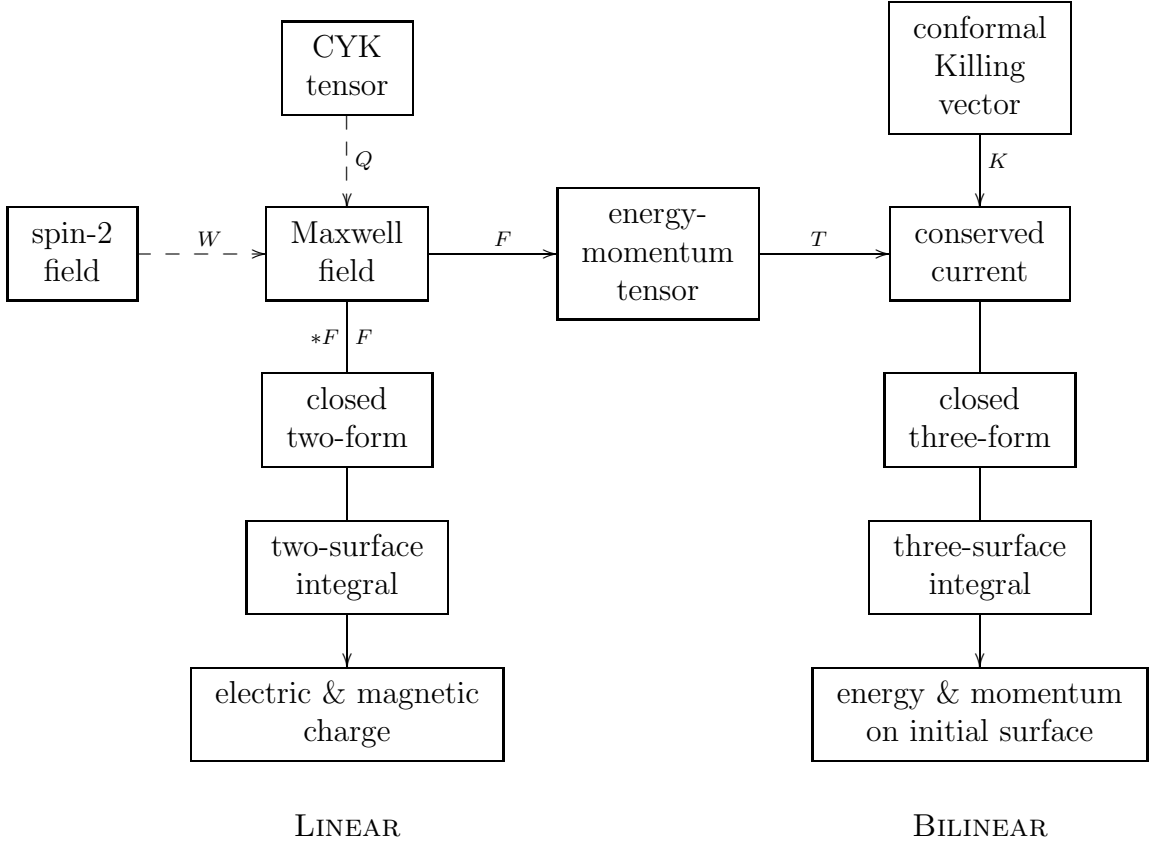
\[
\xymatrix@M=0pt@R=0.7cm@C=0.7cm{
&\fbox{\begin{tabular}{c} CYK \\ tensor \end{tabular}}\ar@{-->}[d]^{Q} & &
 \fbox{\begin{tabular}{c} conformal\\ Killing\\ vector \end{tabular}}\ar[d]^{K}
   \\
 \fbox{\begin{tabular}{c} spin-2 \\ field \end{tabular}}\ar@{-->}[r]^{W}&
  \fbox{\begin{tabular}{c} Maxwell \\ field \end{tabular}}\ar[r]^{F}&
 \fbox{\begin{tabular}{c} energy- \\momentum \\ tensor \end{tabular}}\ar[r]^{T}
 &\fbox{\begin{tabular}{c} conserved \\ current \end{tabular}}    \\
 &
  \fbox{\begin{tabular}{c} closed \\
  two-form \end{tabular}}\ar@{-}[u]_{F}^{\ast F}&
 &\fbox{\begin{tabular}{c} closed \\ three-form \end{tabular}}\ar@{-}[u]    \\
 &
  \fbox{\begin{tabular}{c} two-surface \\ integral \end{tabular}}\ar@{-}[u]\ar[d]&
 &\fbox{\begin{tabular}{c} three-surface \\ integral \end{tabular}}\ar@{-}[u]\ar[d]
    \\
 &
  \fbox{\begin{tabular}{c} electric \& magnetic \\ charge \end{tabular}}\ar@{-}[u]&
 &\fbox{\begin{tabular}{c} energy \& momentum \\ on initial surface
 \end{tabular}}\ar@{-}[u]
    \\
 & \mbox{\sc Linear} & & \mbox{\sc Bilinear}    \\
 } \]

\vspace{0.5cm}

In electrodynamics the linear quantity corresponds to electric or magnetic charge
and the quadratic one expresses the energy, linear momentum or angular momentum of
the Maxwell field. In gravity both kinds of charges play a role of energy.
The linear conserved quantities (as two-surface integrals)
correspond to ADM mass and linear or angular momentum but
bilinear ones are not obviously related to energy. They rather
play a role of energy estimates like in \cite{Ch-Kl} (cf. \cite{LarsAnd}).
In this paper we propose an interpretation of our quadratic
quantity.

We present examples of both kinds of
conserved quantities for a stationary rotating black hole
described by Kerr spacetime.
The linear charge measures (total) ADM mass of black hole in a quasi-local way.
On the other hand, the quadratic functional we interpret as a rotational energy
of the black hole which is not obvious.
Moreover, we derive new conformal (symmetric) Killing tensors which
enable one to construct additional constants of motion along null geodesics.

This paper is organized as follows: In the next Section we call
some basic facts about CYK two-forms in (pseudo-)Riemannian manifolds.
Section 3 contains a small review about CYK tensors and
their applications in four-dimensional spacetime.
In Section \ref{Kerrch} we specify our framework to the case of
a stationary rotating black hole described by the Kerr metric.
Section 5 is devoted to the conserved quantities which are quadratic
in terms of the spin-2 field. One of these charges is applied in Section 6
to the description of rotational energy of the Kerr black hole.
To clarify the exposition some of the technical results and proofs have been
shifted to the appendix.
Moreover, in Appendix B we have placed
 one theorem about CYK tensors on compact Riemannian manifolds.

\section{A short course about CYK tensors}
Let $M$ be an $n$-dimensional ($n>1$) manifold with a Riemannian
or pseudo-Riemannian metric $g_{\mu\nu}$. The covariant derivative
associated with the~Levi--Civita connection will be denoted by
$\nabla$ or just by ``$\,;\,$''. By $T_{...(\mu\nu)...}$ we will
denote the symmetric part and by $T_{...[\mu\nu]...}$ the
skew-symmetric part of tensor $T_{...\mu\nu...}$ with respect to
indices $\mu$ and $\nu$ (analogous symbols will be used for more
indices).

Let $Q_{\mu\nu}$ be a skew-symmetric tensor field (two-form) on
$M$ and let us denote by ${\cal Q}_{\lambda \kappa \sigma}$ a
(three-index) tensor which is defined as follows:
\begin{equation}\label{CYK_eq1}
    {\cal Q}_{\lambda \kappa
    \sigma}(Q,g):= Q_{\lambda \kappa ;\sigma} +Q_{\sigma \kappa
    ;\lambda} - \frac{2}{n-1} \left( g_{\sigma
    \lambda}Q^{\nu}{_{\kappa ;\nu}} + g_{\kappa (\lambda }
    Q_{\sigma)}{^{\mu}}{_{ ;\mu}} \right) \, .
\end{equation}
The object $\cal Q$ has the following algebraic properties
\be\label{wlQ}
   {\cal Q}_{\lambda\kappa\mu}g^{\lambda\mu}=0=
   {\cal Q}_{\lambda\kappa\mu}g^{\lambda\kappa} \, , \quad
{\cal Q}_{\lambda\kappa\mu} =
{\cal Q}_{\mu\kappa\lambda}\, ,
   \ee
i.e. it is traceless and partially symmetric.
\begin{Definition}\label{CYK_df}
    A skew-symmetric tensor $Q_{\mu\nu}$ is a conformal Yano--Killing tensor
    (or simply CYK tensor) for the metric $g$ iff
    \mbox{${\cal Q}_{\lambda \kappa \sigma}(Q,g) = 0$}.
\end{Definition}
In other words, $Q_{\mu\nu}$ is a conformal Yano--Killing tensor if
it fulfils the following equation:
\begin{equation}\label{CYK_eq2}
    Q_{\lambda \kappa ;\sigma} +Q_{\sigma \kappa ;\lambda} =
    \frac{2}{n-1} \left( g_{\sigma \lambda}Q^{\nu}{_{\kappa ;\nu}} +
    g_{\kappa (\lambda } Q_{\sigma)}{^{\mu}}{_{ ;\mu}} \right) \,
\end{equation}
(first proposed by Tachibana and Kashiwada, cf. \cite{Tachibana}).

A more abstract way {with no indices} of describing a CYK
tensor can be found in \cite{BCK}, \cite{Moroianu},
\cite{Semmelmann} or \cite{Stepanow},
where it is considered as the element of the kernel of the twistor operator $Q
\rightarrow {\cal T}_{wist} Q$ defined\footnote{Obviously
${\cal T}_{wist} Q$ corresponds to tensor
${\cal Q}(Q,g)$ (in abstract index notation). Here $X$ is a vector field, $Q$
is a $p$-form, $g: TM \rightarrow T^*M$ is a Riemannian metric, $\rd^*$
denotes coderivative etc.
Conformal Killing $p$-forms are defined with the help of natural differential
operators on Riemannian manifolds. 
We know from the representation theory of the orthogonal group, that the
space of $p$-form valued one-forms ($T^*M \bigotimes\bigwedge^p T^*M$)
decomposes into the orthogonal and
irreducible sum of forms of degree $p+1$ (which gives the exterior
differential $\rd$), the forms of degree $p-1$ (defined by the coderivative
$\rd^*$) and the trace-free part of the partial symmetrization
(the corresponding first order operator is denoted by ${\cal T}_{wist}$).}
as follows:
\[ \forall X \;\; {\cal T}_{wist} Q (X) :=
\nabla_X Q -\frac1{p+1} X \lrcorner \rd Q +
\frac1{n-p+1}  g(X) \wedge \rd^* Q \, .\]
However, to simplify the exposition, we prefer abstract index notation
which also seems to be more popular.

Equation~(\ref{CYK_eq2}) may be transformed into the following
equivalent form:
\begin{equation}\label{CYK_eq3}
    Q_{\lambda (\kappa ;\sigma)} -Q_{\kappa (\lambda ;\sigma)} + \frac
    3{n-1} g_{\sigma[\lambda} Q_{\kappa ]}{^\delta}_{;\delta} =0
\end{equation}
and this is a generalization of the equation
\begin{equation}\label{CYK_eq4}
    Q_{\lambda (\kappa ;\sigma)} -Q_{\kappa (\lambda ;\sigma)} +
    \eta_{\sigma[\lambda} Q_{\kappa ]}{^\delta}_{;\delta} =0 \, ,
\end{equation}
which appeared in \cite{Pen-Rin} as the equation for
skew-symmetric tensor field $Q_{\mu\nu}$ in Minkowski spacetime
with the metric $\eta_{\mu\nu}$. We will show in the sequel that,
 as in \cite{Pen-Rin}, equation~(\ref{CYK_eq2})
enables one to use $Q_{\mu\nu}$ for defining conserved charges
associated with the Weyl tensor.

Using the following symbol:
\begin{equation}\label{div_Q}
    \xi_{\mu}:=Q^{\nu}{}_{\mu;\nu} \, ,
\end{equation}
we can rewrite  equation~(\ref{CYK_eq2}) in the form:

\begin{equation}\label{CYK_eq5}
    Q_{\lambda \kappa ;\sigma} +Q_{\sigma \kappa ;\lambda} =
    \frac{2}{n-1}\left(g_{\sigma\lambda}\xi_{\kappa}
    -g_{\kappa(\lambda}\xi_{\sigma)}\right).
\end{equation}
Let us notice that if $\xi_{\mu}=0$, then $Q_{\mu\nu}$ fulfils the equation:
\begin{equation}\label{Yano_eq}
    Q_{\lambda \kappa ;\sigma} +Q_{\sigma \kappa ;\lambda} = 0 \ .
\end{equation}
Skew-symmetric tensors fulfiling  equation~(\ref{Yano_eq}) are
called in the literature Yano tensors (or Yano--Killing tensors;
see~\cite{Ben-Fran}, \cite{Gib-Rub}, \cite{SKM}, \cite{Yano}). It is obvious
that a two-form $Q_{\mu\nu}$ is a Yano tensor iff $Q_{\mu\nu ;
\lambda}$ is totally skew-symmetric in all indices. So, if
$Q_{\mu\nu}$ fulfils~(\ref{Yano_eq}), then $\xi_{\mu} =
g^{\kappa\lambda}Q_{\kappa\mu ; \lambda} = 0$ (because
$g^{\kappa\lambda}$ is symmetric in its indices). That means that
each Yano tensor is a conformal Yano--Killing tensor, but not the
other way around. The necessary and sufficient condition for a CYK
tensor to be a Yano tensor is the vanishing of~$\xi_{\mu}$.

CYK tensors are the conformal generalization of Yano tensors. More
precisely, for any positive function $\Omega$ on $M$ the tensor
$\cal Q$ 
transforms with respect to the
conformal rescaling as follows:
\begin{equation}\label{conf_resc}
    {\cal Q}_{\lambda \kappa \sigma} (Q,g) = \Omega^{-3}
    {\cal Q}_{\lambda \kappa \sigma} (\Omega^3 Q,\Omega^2 g) \, ,
\end{equation}
which implies the following
\begin{Theorem}\label{conf_resc_th}
    If $Q_{\mu\nu}$ is a CYK tensor for the metric $g_{\mu\nu}$,
    then $\Omega^3 Q_{\mu\nu}$ is a CYK tensor for the conformally rescaled metric
    $\Omega^2 g_{\mu\nu}$.
\end{Theorem}
Proof of the formula (\ref{conf_resc}) is given in Appendix~\ref{conf_resc_ch}.
The form of equation~(\ref{Yano_eq}) does not remain unchanged under
conformal rescaling. But that equation is a particular case of
equation (\ref{CYK_eq2}) whose form remains unchanged under such a
transformation. That means that if $Q$ is a Yano tensor of the
metric $g$, then although in general $\Omega^3 Q$ is not a Yano
tensor of the metric $\Omega^2 g$, it is a CYK tensor. In this
sense equation~(\ref{CYK_eq2}) is a conformal generalization of
 equation~(\ref{Yano_eq}).

\subsection{The connection between CYK tensors and Killing tensors}

It is a known fact that the ``square'' of a Yano tensor is a Killing tensor. It turns out
that in the same way CYK tensors are connected with conformal Killing tensors.
In the following definitions we will restrict ourselves to the Killing tensors
and conformal Killing tensors of rank 2, although one can consider tensors
of any rank (\cite{ColHow}, \cite{SKM}).

\begin{Definition}
    A symmetric tensor $A_{\mu\nu}$ is a Killing tensor iff it fulfils
    the equation:
    \begin{equation}\label{K_tens_eq}
        A_{(\mu\nu ; \kappa)} = 0.
    \end{equation}

\end{Definition}

\begin{Definition}
    A symmetric tensor $A_{\mu\nu}$ is a conformal Killing tensor iff
    it fulfils the equation:
    \begin{equation}\label{CK_tens_eq}
        A_{(\mu\nu ; \kappa)} = g_{(\mu\nu}A_{\kappa)}
    \end{equation}
    for a certain covector $A_{\kappa}$.
\end{Definition}
It is obvious that equation~(\ref{K_tens_eq}) is a particular case of
(\ref{CK_tens_eq}). It is easy to see that the covector~$A_{\kappa}$ is unambiguously
determined by  equation~(\ref{CK_tens_eq})
(it can be shown e.g. by contracting
the equation with~$g^{\mu\nu}$). To be more precise:
\[
A_{\kappa} = \frac{1}{n+2}(2A^{\mu}{}_{\kappa ; \mu} +
A^{\mu}{}_{\mu ; \kappa}).\]
From the above definitions one can
easily see that a (conformal) Killing tensor is a generalization of
a (conformal) Killing vector (cf. \cite{Carter},  \cite{HS},
\cite{REB}, \cite{Pen-Walk}, \cite{Woodhouse}).

Obviously, if $Q_{\mu\nu}$ is a skew-symmetric tensor,
then $A_{\mu\nu}$ defined by the formula
\begin{equation}\label{Q_sqr}
    A_{\mu\nu} = Q_{\mu\lambda}Q^{\lambda}{}_{\nu}
\end{equation}
is a symmetric tensor. It turns out that if a skew-symmetric tensor $Q_{\mu\nu}$
fulfils  equation~(\ref{CYK_eq2}) (it is a CYK tensor), then
$A_{\mu\nu}$ defined by~(\ref{Q_sqr})
fulfils  equation~(\ref{CK_tens_eq}) with
\begin{equation}\label{A_vect}
    A_{\kappa} = \frac{2}{n-1} Q_{\kappa}{}^{\lambda}Q_{\lambda}{}^{\delta}{}_{;\delta}
\end{equation}
(therefore -- since $A_{\mu\nu}$ is symmetric -- it is a conformal
Killing tensor). If $Q_{\mu\nu}$ is a Yano tensor, then
$A_{\kappa}$  defined by formula~(\ref{A_vect}) vanishes, thus
$A_{\mu\nu}$ (defined by~(\ref{Q_sqr})) is a Killing tensor. That
enables one to formulate the following (cf. Prop. 5.1. in \cite{Gib-Rub} or
35.44 in \cite{SKM})
\begin{Theorem}\label{killing_tens_th}
If \ $Q_{\mu\nu}$ and $P_{\mu\nu}$ are (conformal) Yano--Killing tensors,
then the symmetrized product $A_{\mu\nu} := Q_{\lambda(\mu}P_{\nu)}{^{\lambda}}$ is
a (conformal) Killing tensor.
\end{Theorem}
\begin{proof} 
Let $Q_{\mu\nu}$ and $P_{\mu\nu}$ be conformal Yano-Killing tensors. We have then
\[
Q_{\kappa \lambda ;\sigma} +Q_{\sigma \lambda ;\kappa} =   \frac{2}{n-1}\left(g_{\sigma\kappa}\xi_{\lambda}-g_{\lambda(\kappa}\xi_{\sigma)}\right)
\]
and
\[
P_{\kappa \lambda ;\sigma} + P_{\sigma \lambda ;\kappa}
= \frac{2}{n-1}\left(g_{\sigma\kappa}\zeta_{\lambda}-g_{\lambda(\kappa}\zeta_{\sigma)}\right),
\]
where $\xi_{\mu}=Q^{\nu}{}_{\mu;\nu}$ and $\zeta_{\mu}=P^{\nu}{}_{\mu;\nu}$.
Contracting the first of the above equations with $P_{\nu}{}^{\lambda}$, we get
\[
Q_{\kappa \lambda ;\sigma}P_{\nu}{}^{\lambda}
+ Q_{\sigma \lambda ;\kappa}P_{\nu}{}^{\lambda} =
  \frac{2}{n-1}\left(g_{\sigma\kappa}\xi_{\lambda}P_{\nu}{}^{\lambda}
  - \frac12 P_{\nu\kappa}\xi_{\sigma} - \frac12 P_{\nu\sigma}\xi_{\kappa}\right).
\]
Symmetrizing this equation in $\kappa$, $\sigma$ and $\nu$,
we get (since $P$ is skew-symmetric)
\[
Q_{\lambda (\kappa ;\sigma}P_{\nu)}{}^{\lambda} =
-\frac{1}{n-1} g_{(\kappa\sigma} P_{\nu)}{}^{\lambda}\xi_{\lambda}.
\]
Analogously we get
\[
Q_{\lambda (\kappa}P_{\nu}{}^{\lambda}{}_{;\sigma)} =
- \frac{1}{n-1} g_{(\kappa\sigma} Q_{\nu)}{}^{\lambda}\zeta_{\lambda}.
\]
Finally, if $A_{\kappa\nu}:=Q_{\lambda(\kappa}P_{\nu)}{}^{\lambda}$, then
\begin{eqnarray}
A_{(\kappa\nu;\sigma)} & = &
\left( Q_{\lambda(\kappa}P_{\nu}{}^{\lambda} \right){_{; \sigma )}} =
Q_{\lambda (\kappa ;\sigma}P_{\nu)}{}^{\lambda}
+ Q_{\lambda (\kappa}P_{\nu}{}^{\lambda}{}_{;\sigma)} \nonumber \\
 & = & -\frac{1}{n-1} \left( g_{(\kappa\sigma} P_{\nu)}{}^{\lambda}\xi_{\lambda}
+ g_{(\kappa\sigma} Q_{\nu)}{}^{\lambda}\zeta_{\lambda} \right) =
g_{(\kappa\nu} A_{\sigma)}, \nonumber
\end{eqnarray}
where
\[
A_{\nu} := \frac{1}{n-1} \left( P^{\lambda}{}_{\nu}\xi_{\lambda}
+ Q^{\lambda}{}_{\nu}\zeta_{\lambda} \right).
\]
This means that $A_{\kappa\nu}$ is a conformal Killing tensor.
If now $Q$ and $P$ are Yano tensors, then $\xi_{\mu}=\zeta_{\mu}=0$,
which implies $A_{\mu}=0$.  In that case $A_{\kappa\nu}$
is a Killing tensor.

\end{proof}

\subsection{The connection between CYK tensors and Killing vectors}
Let us denote by $R^{\sigma}{}_{\kappa\lambda\mu}$ the Riemann
tensor describing the curvature of the manifold $(M,g)$. We use now an
integrability condition {\setlength\arraycolsep{2pt}
\begin{eqnarray}\label{war_calk1}
    2Q_{\lambda\kappa;\nu\mu} & = &
    \frac{2}{n-1}\left(g_{\lambda\mu}\xi_{\kappa;\nu}
    + g_{\nu\lambda}\xi_{\kappa;\mu} - g_{\mu\nu}\xi_{\kappa;\lambda}
    - g_{\kappa(\lambda}\xi_{\mu);\nu}
    + g_{\kappa(\mu}\xi_{\nu);\lambda} -
    g_{\kappa(\nu}\xi_{\lambda);\mu}\right)\nonumber\\[2pt] & &
    + Q_{\sigma\lambda}R^{\sigma}{}_{\kappa\mu\nu}
    + Q_{\sigma\mu}R^{\sigma}{}_{\kappa\lambda\nu}
    + Q_{\sigma\nu}R^{\sigma}{}_{\kappa\lambda\mu}
    + 2Q_{\sigma\kappa}R^{\sigma}{}_{\mu\nu\lambda} \, ,
\end{eqnarray}}
 which we prove in Appendix~\ref{war_calk_dod}.

 A contraction in indices $\kappa$ and $\nu$ gives us:
\begin{equation}\label{zzz}
    g_{\mu\lambda}\xi^{\sigma}{}_{;\sigma} + (n-2)\xi_{(\mu;\lambda)} =
    (n-1)R_{\sigma(\mu} Q_{\lambda)}{}^{\sigma} \, ,
\end{equation}
where by $R_{\mu\nu}$ we denote the Ricci tensor of the metric
$g_{\mu\nu}$. Taking the trace of \eq{zzz} we obtain:
\[ (2n-2)\xi^{\sigma}{}_{;\sigma} = (n-1)R_{\sigma\mu}Q^{\mu\sigma} = 0 \ ,\]
where the last equality results from the fact that~$R_{\sigma\mu}$
is a symmetric tensor and~$Q^{\mu\sigma}$ is a skew-symmetric one.
Therefore,  we have $\xi^{\sigma}{}_{;\sigma} = 0$ which, for $n>2$,
implies
\begin{equation}\label{almost_killing}
    \xi_{(\mu;\lambda)}=\frac{n-1}{n-2}R_{\sigma(\mu}Q_{\lambda)}{}^{\sigma}.
\end{equation}

If $M$ is an Einstein manifold,
i.e. its Ricci tensor~$R_{\mu\nu}$ is proportional to its metric
$g_{\mu\nu}$, then using equation~(\ref{almost_killing}) we obtain
\[
    \xi_{(\mu;\nu)}=-\frac{n-1}{n-2}\lambda g_{\sigma(\mu}Q_{\nu)}{}^{\sigma}
    = -\frac{n-1}{n-2}\lambda Q_{(\nu\mu)} = 0.
\]
Here $R_{\mu\nu}= - \lambda g_{\mu\nu}$ and by $\lambda$ we denote
a cosmological constant. The condition $\xi_{(\mu;\lambda)}=0$ means
that $\xi^{\mu}$ is a Killing vector field of the
metric~$g_{\mu\nu}$. That enables one to formulate the following
\begin{Theorem}\label{killing_th}
If $g_{\mu\nu}$ is a solution of the vacuum Einstein equations with
cosmological constant and $Q_{\mu\nu}$ is its CYK tensor, then
$\xi^{\mu}=\nabla_\nu Q^{\nu\mu}$ is a Killing vector field of the
metric~$g_{\mu\nu}$.
\end{Theorem}
Let us notice that this fact reduces the number of Einstein
metrics possessing a nontrivial CYK tensor. The existence of a
solution of equation~(\ref{CYK_eq2}) 
which is \emph{not} a Yano tensor implies that our manifold $M$ has
at least one symmetry. In the case of a Yano tensor that does not have to
be true.

\section{CYK tensors in four dimensions}

In the following Section we restrict ourselves to an oriented
manifold $M$ with dimension $n = 4$. It turns out that in this
case conformal Yano--Killing tensors possess some additional
properties.

\subsection{Hodge duality}
In the space of differential forms of an oriented manifold one can
define a mapping called a Hodge duality (Hodge star). It assigns
to every $p$--form a $(n-p)$--form ($n$ is the dimension of the
manifold). We consider the case of $n=4$ and $p=2$. The Hodge star then
becomes  a mapping which assigns to a two-form $\omega$ a
two-form $*\omega$. Using coordinates we can express this mapping
in the following way:
\begin{equation}\label{Hodge_star}
    *\omega_{\alpha\beta} :=
    \frac{1}{2}\varepsilon_{\alpha\beta}{}^{\mu\nu}\omega_{\mu\nu} \, ,
\end{equation}
where $\varepsilon_{\alpha\beta\mu\nu}$ is the skew-symmetric
Levi--Civita tensor\footnote{It can be defined by the formula
$\varepsilon_{\alpha\beta\mu\nu}= \sqrt{|\det g|}
\epsilon_{\alpha\beta\mu\nu}$, where
\[\epsilon_{\alpha\beta\mu\nu} =
\left\{ \begin{array}{ll}
+1 & \textrm{if $\alpha\beta\mu\nu$ is an even permutation of $0, 1, 2, 3$}\\
-1 & \textrm{if $\alpha\beta\mu\nu$ is an odd permutation of $0, 1, 2, 3$}\\
\ \ \, 0 & \textrm{in any other case}
\end{array}\right.\]}
determining the orientation of the manifold
($\frac{1}{4!}\varepsilon_{\alpha\beta\mu\nu} \rd x^{\alpha}
\wedge \rd x^{\beta} \wedge \rd x^{\mu} \wedge \rd x^{\nu}$ is the
volume form of the manifold $M$). For Riemannian metrics (with the
positive signature) we have $**\omega = \omega$ while for
Lorentzian metrics the negative sign appears $**\omega = -\omega$
(we keep assuming that $\omega$ is a two-form defined on a
four-dimensional manifold $M$). The last equality shows that in
the Lorentzian case we cannot have $*\omega=\pm \omega$, whereas for
positive metrics it is possible. If $*\omega = \omega$ ($*\omega =
- \omega$), we say that $\omega$ is selfdual (anti-selfdual).

Due to a CYK tensor being a two-form, it is reasonable to ask what
are the properties of its dual. Let $Q$ be a conformal
Yano--Killing tensor and~$*Q$ its dual. Moreover, 
similarly to $\xi_{\mu}=\nabla^{\nu}Q_{\nu\mu}$ (cf. \eq{div_Q})
let us introduce the following covector
\begin{equation}\label{div_QQstar}
    \chi_{\mu} := \nabla^{\nu} \ast\! Q_{\nu\mu} \, .
\end{equation}
Contracting \eq{div_QQstar} with $\frac{1}{2}\varepsilon_{\alpha
\beta}{}^{\lambda\kappa}$ we obtain 
\begin{equation}\label{Qstar_der}
\nabla_{\sigma} \ast\! Q_{\alpha\beta} =
\frac{2}{3}g_{\sigma[\alpha}\chi_{\beta] } + \frac{1}{3}
\varepsilon_{\alpha\beta\sigma\kappa}\xi^{\kappa}\, .
\end{equation}
If we multiply  equation~(\ref{Qstar_der}) by $\frac{1}{2}\varepsilon_{\mu\nu}
{}^{\alpha\beta}$, we obtain the analogous equality:

\begin{equation}\label{Q_der}
    Q_{\mu\nu;\sigma}=\frac{2}{3}g_{\sigma[\mu}\xi_{\nu]}\pm \frac{1}{3}
    \varepsilon_{\mu\nu\sigma\beta}\chi^{\beta}\ .
\end{equation}
The plus sign in  equation~(\ref{Q_der}) refers to metrics with
the positive determinant and the minus sign refers to metrics with
the negative one.

If we symmetrize  equation (\ref{Qstar_der}) in indices
$\alpha$ and $\sigma$, we receive the following equality:
\begin{equation}\label{Qstar_CYK}
\ast Q_{\alpha\beta;\sigma}+\ast Q_{\sigma\beta;\alpha} =
\frac{2}{3}\left(g_{\sigma\alpha}
    \chi_{\beta}-g_{\beta(\alpha}\chi_{\sigma)}\right)\, .
\end{equation}
It is not hard to recognize that this is  equation~(\ref{CYK_eq2})
for the tensor $*Q$. It proves the following
\begin{Theorem}\label{dual_th}
Let $g_{\mu\nu}$ be a metric of a four-dimensional differential
manifold $M$. A skew-symmetric tensor $Q_{\mu\nu}$ is a CYK tensor
of the metric $g_{\mu\nu}$ iff its dual $*Q_{\mu\nu}$ is a CYK
tensor of this metric.
\end{Theorem}
The above theorem implies that for every four-dimensional manifold
solutions of  equation (\ref{CYK_eq2}) appear in pairs -- to
each solution we can assign the solution dual to it (in the Hodge
duality sense). Now we can observe that  equation~(\ref{Q_der})
is  equation~(\ref{Qstar_der}), in which $Q$ was replaced by
$*Q$ ($\pm$ in  equation~(\ref{Q_der}) reflects the fact that
$*\!*\!Q=\pm Q$).

Theorems \ref{killing_th} and \ref{dual_th} imply that an Einstein
metric for which there exists a solution of  equation~(\ref{CYK_eq2})
 has in general two symmetries. However,
it does not have to be so if at least one of the fields
$\xi^{\mu}$ or $\chi^{\mu}$ is equal to zero (i.e. at least one of
the CYK tensors is a Yano tensor) or if those fields are not
linearly independent ($\chi^{\mu} = c\, \xi^{\mu}, \
c=\textrm{const.}$).

\subsection{Spin-2 field}\label{subspin2}
In order to be able to use conformal Yano--Killing tensors for the
construction of charges in general relativity we have to introduce
the notion of a spin-2 field. For the purpose of this paper we
introduce the following definition.

\begin{Definition}\label{spin2_df}
    The tensor field $W_{\alpha\beta\mu\nu}$ is called a spin-2 field iff
    the following equations are fulfiled:
    \[W_{\alpha\beta\mu\nu} = W_{\mu\nu\alpha\beta} = W_{[\alpha\beta][\mu\nu]},\]
    \begin{equation}\label{spin2_prop}
        W_{\alpha[\beta\mu\nu]} = 0,\ \ g^{\alpha\mu}W_{\alpha\beta\mu\nu} = 0
    \end{equation}
    and
    \begin{equation}\label{Bianchi_id}
        \nabla_{[\lambda} W_{\mu\nu ] \alpha\beta} =0.
    \end{equation}
\end{Definition}
According to the above definition the spin-2 field is any Riemann
tensor of a vacuum metric (i.e. Ricci flat metric), as well as any
Weyl tensor of linearized Einstein theory.
Equations~(\ref{Bianchi_id}) (which in the case
of~$W_{\mu\nu\alpha\beta}$ being a Riemann tensor are simply Bianchi
identities) can be treated as the field equations for
$W_{\mu\nu\alpha\beta}$.

$W_{\mu\nu\alpha\beta}$ is skew-symmetric both in the first and
the second pair of indices and to both of them the Hodge star can
be applied. We denote
\[
{}^{*}W_{\mu\nu\alpha\beta} = \frac{1}{2}\varepsilon_{\mu\nu}{}^{\rho\sigma}
W_{\rho\sigma\alpha\beta} \ ,\ \ \ \ \ \ W^{*}{}_{\mu\nu\alpha\beta} =
\frac{1}{2}W_{\mu\nu\rho\sigma}\varepsilon^{\rho\sigma}{}_{\alpha\beta}\ .
\]
One can show that
\[
{}^{*}W = W^{*}, \ \ \ \ \ \ {}^{*}({}^{*}W) = {}^{*}W^{*} = \pm W.
\]
The ``$+$'' sign in the above equation refers to (positive)
Riemannian metrics and the ``$-$'' sign refers to Lorentzian
metrics. One can also show that if $W$ is a spin-2 field, then
${}^{*}W$ (and consequently $W^{*}$) is a spin-2 field as well.
If we assume that $W$ fulfils  equations~(\ref{spin2_prop}),
then the following equivalences are satisfied:
\begin{eqnarray}\label{Bianchi_equiv}
\nabla_{[\lambda}W_{\mu\nu]\alpha\beta} = 0 \iff \nabla^{\mu}W_{\mu\nu\alpha\beta} = 0 \iff
\nonumber \\
\iff \nabla_{[\lambda}{}^{*}W_{\mu\nu]\alpha\beta} = 0
\iff \nabla^{\mu}{}^{*}W_{\mu\nu\alpha\beta} = 0 \ .
\end{eqnarray}

\subsection{The relation between CYK tensors and spin-2 fields}
Let $W_{\mu\nu\alpha\beta}$ be a spin-2 field and $Q_{\mu\nu}$
be any skew-symmetric tensor. Let us denote by $F$ the following
two-form
\begin{equation}\label{def_FWQ}
    F_{\mu\nu}(W,Q) := W_{\mu\nu\lambda\kappa}Q^{\lambda\kappa}.
\end{equation}
The following formula is satisfied:
\begin{equation}\label{divF}
    \nabla_{\nu}F^{\mu\nu}(W,Q) = \frac{2}{3} W^{\mu\nu\alpha\beta}
{\cal Q}_{\alpha\beta\nu}(Q,g)\
\end{equation}
which may be argued as follows:\\
\begin{proof} Let us notice firstly that due to $W_{\mu\nu\alpha\beta}$
being traceless we have:
{\setlength\arraycolsep{2pt}
\begin{eqnarray}\label{WQ3}
    W^{\mu\nu\alpha\beta}{\cal Q}_{\alpha\beta\nu} & = & W^{\mu\nu\alpha\beta}
    (Q_{\alpha\beta;\nu} + Q_{\nu\beta;\alpha}) = (W^{\mu\nu\alpha\beta}
    + W^{\mu\alpha\nu\beta}) Q_{\alpha\beta;\nu} \nonumber\\
    & = & (W^{\mu\nu\alpha\beta} + \frac{1}{2} W^{\mu\alpha\nu\beta}
    - \frac{1}{2} W^{\mu\beta\nu\alpha}) Q_{\alpha\beta;\nu} \nonumber\\
    & = & \frac{3}{2} W^{\mu\nu\alpha\beta} Q_{\alpha\beta;\nu} \ ,
\end{eqnarray}}\noindent
where we have used the fact that $Q_{\alpha\beta;\nu}$ is
skew-symmetric in indices $\alpha$ and $\beta$ and
$W^{\mu[\nu\alpha\beta]} = 0$. Using the above formula~(\ref{WQ3}) and
properties (\ref{Bianchi_equiv}) of a spin-2 field
$W_{\mu\nu\alpha\beta}$ we obtain the final result 
{\setlength\arraycolsep{2pt}
\begin{eqnarray}
\nabla_{\nu}F^{\mu\nu}(W,Q) & = & \nabla_{\nu} \left(
W^{\mu\nu\alpha\beta}Q_{\alpha\beta} \right) =
(\nabla_{\nu}W^{\mu\nu\alpha\beta})Q_{\alpha\beta}  +
W^{\mu\nu\alpha\beta}(\nabla_{\nu}Q_{\alpha\beta})
\nonumber \\
& = & \frac{2}{3} W^{\mu\nu\alpha\beta} {\cal Q}_{\alpha\beta\nu}
\ .\nonumber
\end{eqnarray}}
\end{proof}
Definition~\ref{CYK_df} means that ${\cal
Q}_{\alpha\beta\nu}(Q,g)=0$, hence for a CYK tensor $Q_{\mu\nu}$
we have
\begin{equation}\label{divF_CYK}
\nabla_{\nu}F^{\mu\nu}(W,Q) = 0 \ .
\end{equation}
Let $V$ be a three-volume and $\partial V$ its boundary. Formula
 \eq{divF_CYK} implies\footnote{Symbols $\rd \sigma_{\mu\nu}$
 and $\rd \Sigma_{\mu}$
can be defined in the following way: if $\Omega$ stands for the volume
form of the manifold $M$, then $\rd \sigma_{\mu\nu}:=(\partial_\mu
\wedge \partial_\nu) \lrcorner\, \Omega$, $\rd \Sigma_{\mu}:=
\partial_\mu \lrcorner\, \Omega$.}
\[
\int_{\partial V} F^{\mu\nu}(W,Q) \rd \sigma_{\mu\nu} = \int_{V}
\nabla_{\nu} F^{\mu\nu}(W,Q) \rd \Sigma_{\mu} = 0 \, ,
\]
which means that the flux of $F^{\mu\nu}$ through any
two-dimensional closed surfaces $S_{1}$ and $S_{2}$ is the same as
long as we are able to find a three-volume $V$ between them (i.e. there
exists $V$ such that $\partial V = S_{1} \cup S_{2}$). In this
sense $Q_{\mu\nu}$ defines a charge related to the spin-2
field\footnote{Equation (\ref{divF}) can be also used to
define asymptotic charges of an asymptotically flat space. Let
$S_t$ be the family of closed two-dimensional surfaces going to
spatial infinity with $t\to \infty$. The formula $\lim_{t\to
\infty}\int_{S_t} F^{\mu\nu}(Q,g) \rd \sigma_{\mu\nu}$ defines a
quantity (an asymptotic charge) which does not depend on the
choice of the family $S_t$ as long as ${\cal Q}_{\mu\nu\rho}(Q,g)$
goes to zero in spatial infinity fast
enough~(see~\cite{Goldberg1}, \cite{JJspin2}, \cite{cykem}).} $W$.

We shall prove now the following
\begin{Theorem}\label{Maxwell_th}
Let $W$ be a spin-2 field and $Q$ be a conformal Yano--Killing
tensor, then a skew-symmetric tensor $F_{\mu\nu}$ defined by the
formula~(\ref{def_FWQ}) satisfies the vacuum Maxwell equations, i.e.
\begin{equation}\label{Maxwell_eq}
\nabla_{\lambda} F^{\mu\lambda}=0=\nabla_{\lambda} *\! F^{\mu\lambda} \, ,
\end{equation}
where $\ast F^{\mu\lambda}=
\frac12\varepsilon^{\mu\lambda}{}_{\rho\sigma}F^{\rho\sigma}$.
\end{Theorem}
\begin{proof}
The first of  equations~(\ref{Maxwell_eq}) is simply  equation~(\ref{divF}) (since ${\cal Q}_{\alpha\beta\nu}(Q,g)=0$
for $Q$ being a CYK tensor). The second equation is a simple
consequence of the fact that
$*F(W,Q)=F({}^{*}W,Q)=F(W^{*},Q)=F(W,*Q)$ and that $*Q$ is a CYK
tensor. \end{proof}

 Theorem~\ref{Maxwell_th} implies that a charge
defined by the surface integral $\int_{S} F^{\mu\nu} \rd
\sigma_{\mu\nu}$ is the ``electromagnetic'' charge of the field
$F^{\mu\nu}$. More precisely,
due to the Levi--Civita connection being torsionless, we have
\[
\nabla_{[\lambda}F_{\kappa\rho]}=
\partial_{[\lambda}F_{\kappa\rho]}\, .
\]
Moreover,
\begin{equation}\label{Maxwell_2}
0=\nabla_{\lambda} *\! F^{\mu\lambda}=\nabla_{\lambda}(\frac12\varepsilon^{\mu\lambda
\kappa\rho}F_{\kappa\rho})=\frac12\varepsilon^{\mu\lambda\kappa\rho}F_{\kappa\rho;\lambda}=
\frac12\varepsilon^{\mu\lambda\kappa\rho}F_{\kappa\rho,\lambda}\, .
\end{equation}
Multiplying the above equality by
$\varepsilon_{\mu\alpha\beta\gamma}$ we get
$F_{[\alpha\beta,\gamma]}=0$, which simply means that $\rd F=0$.
The formula (\ref{Maxwell_2}) obviously implies that if $\rd F=0$,
then $\nabla_{\lambda} *\! F^{\mu\lambda}=0$. This means that the
equalities~(\ref{Maxwell_eq}) are equivalent to the following
equations:
\begin{equation}\label{Maxwell_eq2}
    \rd F = 0 = \rd *\!F \, .
\end{equation}

\subsection{CYK tensors in Minkowski spacetime}\label{CYK_in_Mink_ch}
Let $M$ be the Minkowski space and $(x^{\mu})$ Cartesian
coordinates on $M$. We have then $g_{\mu\nu}=\eta_{\mu\nu}:=
\textrm{diag}(-1,1,1,1)$. Taking into account that the Riemann
tensor of the metric $\eta_{\mu\nu}$ vanishes
and, in coordinates ${x^{\mu}}$, covariant derivative is the same as
ordinary derivative (we change ,,$\, ; \,$'' into ,,$\,
,\,$''),  equation~(\ref{war_calk}) 
takes the following form:
\be\label{war_calk_Mink}
    Q_{\lambda\kappa,\nu\mu}  =
    \frac1{3}\Big(\eta_{\lambda\mu}\xi_{\kappa,\nu}
    + \eta_{\nu\lambda}\xi_{\kappa,\mu} - \eta_{\mu\nu}\xi_{\kappa,\lambda} -
    \eta_{\kappa(\lambda}\xi_{\mu),\nu}
    + \eta_{\kappa(\mu}\xi_{\nu),\lambda}
    - \eta_{\kappa(\nu}\xi_{\lambda),\mu}\Big) \, ,
    \ee
where $Q_{\mu\nu}$ is a CYK tensor on $M$ and
$\xi^{\nu}=\partial_{\mu}Q^{\mu\nu}$. According to
Theorem~\ref{killing_th}, vanishing of the Ricci tensor implies
that $\xi^{\mu}$ is a Killing vector of the metric
 $\eta_{\mu\nu}$. It is well known that in the Minkowski spacetime
the space of Killing fields is spanned by the fields
\begin{equation}\label{kvf_generators}
{\cal T}_\mu := \frac{\partial}{\partial x^{\mu}},\quad
{\cal L}_{\mu\nu} := x_{\mu}\frac{\partial}{\partial x^{\nu}} -
x_{\nu}\frac{\partial}{\partial x^{\mu}}
\end{equation}
(where $x_{\mu}=\eta_{\mu\nu}x^{\nu}$). It means that components of the vector $\xi^{\mu}$
in coordinates $(x^{\mu})$ are polynomials of degree one in variables $x^{\mu}$.
Equation~(\ref{war_calk_Mink}) implies that coordinates of the tensor $Q_{\mu\nu}$
also have to be polynomials of degree at most two. We have then
$Q_{\mu\nu}=Q^{\scriptscriptstyle(0)}{}_{\mu\nu} + Q^{\scriptscriptstyle(1)}
{}_{\mu\nu} + Q^{\scriptscriptstyle(2)}{}_{\mu\nu}$, where
$Q^{\scriptscriptstyle(i)}{}_{\mu\nu}$ is a monomial of degree $i$.
If we use this decomposition in  equation~(\ref{CYK_eq2}), it becomes obvious
that each tensor~$Q^{\scriptscriptstyle(i)}{}_{\mu\nu}$ has to fulfil it separately.
Let us see what conditions it imposes on those monomials.

First, it is easy to see that every constant skew-symmetric tensor fulfils  equation~(\ref{CYK_eq2}) and therefore $Q^{\scriptscriptstyle(0)}{}_{\mu\nu}$
stays undetermined. Let us focus then on~$Q^{\scriptscriptstyle(1)}{}_{\mu\nu}$.
We can write $Q^{\scriptscriptstyle(1)}{}_{\mu\nu}=A_{\mu\nu\sigma}x^{\sigma}$,
where $A_{\mu\nu\sigma}$ is a certain constant tensor (skew-symmetric in indices
$\mu$ and $\nu$). $Q^{\scriptscriptstyle(1)}{}_{\mu\nu}$ is linear in $x^{\mu}$
and therefore $\xi^{\scriptscriptstyle(1)}{}^{\nu}:=Q^{\scriptscriptstyle(1)}
{}^{\mu\nu}{}_{,\mu}$ cannot depend on $x^{\mu}$. We have then
$\xi^{\scriptscriptstyle(1)}{}^{\mu}=u^{\mu}$, where $u^{\mu}$ denotes a certain
constant vector. The equation $Q^{\scriptscriptstyle(1)}{}^{\mu\nu}{}_{,\mu}=u^{\nu}$
reduces to~$\eta^{\mu\sigma}A_{\mu\nu\sigma}=u_{\nu}$. Decomposing $A_{\mu\nu\sigma}$
as
$A_{\mu\nu\sigma}=\frac23\eta_{\sigma[\mu}u_{\nu]} + B_{\mu\nu\sigma}$ we get
$B_{\mu\nu\sigma}=B_{[\mu\nu]\sigma}$ and $\eta^{\mu\sigma}B_{\mu\nu\sigma}=0$.
Simple calculation shows that \mbox{$Q^{\scriptscriptstyle(1')}{}_{\mu\nu}:=
\frac23\eta_{\sigma[\mu}u_{\nu]}x^{\sigma}=\frac23x_{[\mu}u_{\nu]}$} satisfies
 equation~(\ref{CYK_eq2}), which means that tensor
\mbox{$Q^{\scriptscriptstyle(1'')}
{}_{\mu\nu}:= B_{\mu\nu\sigma}x^{\sigma}$} also satisfies it. Yet
$Q^{\scriptscriptstyle(1'')}{}^{\mu\nu}{}_{,\mu}:=
\eta^{\mu\sigma}B_{\mu\nu\sigma}=0$,
which means that $Q^{\scriptscriptstyle(1'')}{}_{\mu\nu,\sigma} +
Q^{\scriptscriptstyle(1'')}{}_{\sigma\nu,\mu} =
B_{\mu\nu\sigma} + B_{\sigma\nu\mu} = 0$.
The last equation is equivalent to the statement that $B_{\mu\nu\sigma}
=B_{[\mu\nu\sigma]}$.
Since $Q^{\scriptscriptstyle(1'')}{}_{\mu\nu}$ is a CYK tensor linear
in $x^{\mu}$,
$*Q^{\scriptscriptstyle(1'')}{}_{\mu\nu}$ also has this property, and
what follows
$*Q^{\scriptscriptstyle(1'')}{}^{\mu\nu}{}_{,\mu}=v^{\nu}$ for
a certain constant vector
$v^{\nu}$. The last equation can be written
as $\frac12\varepsilon^{\sigma\rho\mu\nu}
B_{\mu\nu\sigma}=v^{\rho}$. Multiplying it
by $\varepsilon_{\rho\alpha\beta\gamma}$
we get \mbox{$\varepsilon_{\rho\alpha\beta\gamma}v^{\rho}
=\frac12\delta^{\sigma\mu\nu}
{}_{\alpha\beta\gamma}B_{\mu\nu\sigma}=3B_{[\alpha\beta\gamma]}
=3B_{\alpha\beta\gamma}$}.
Therefore $Q^{\scriptscriptstyle(1'')}{}_{\mu\nu}=
\frac13\varepsilon_{\mu\nu\rho\sigma} v^{\rho}x^{\sigma}$. Finally
$Q^{\scriptscriptstyle(1)}{}_{\mu\nu} = \frac23x_{[\mu}u_{\nu]} +
\frac13\varepsilon_{\mu\nu\rho\sigma}v^{\rho}x^{\sigma}$, where $u^{\mu}$ and
$v^{\mu}$ are any constant vectors. It is easy to check that every tensor
 of this form fulfils  equation~(\ref{CYK_eq2}).

The last thing is computing $Q^{\scriptscriptstyle(2)}{}_{\mu\nu}$. We observe
first that the components of the vector $\xi^{\scriptscriptstyle(2)}{}^{\nu}:=
Q^{\scriptscriptstyle(2)}{}^{\mu\nu}{}_{,\mu}$ in the coordinates $(x^{\mu})$
are monomials of degree one in variables $x^{\mu}$. All Killing fields
with this property have the form $k^{\mu\nu}{\cal L}_{\mu\nu}=
k^{\mu\nu}(x_{\mu}\partial_{\nu}-x_{\nu}\partial_{\mu})=2k_{\mu}{}^{\nu}
x^{\mu}\partial_{\nu}$, where $k^{\mu\nu}$ is a certain constant skew-symmetric tensor.
Then let $k^{\mu\nu}$ be such that $\xi^{\scriptscriptstyle(2)}{}_{\mu}=
k_{\mu\nu}x^{\nu}$. $Q^{\scriptscriptstyle(2)}{}_{\mu\nu}$ can be always written in
the form $Q^{\scriptscriptstyle(2)}{}_{\mu\nu}= \frac12C_{\mu\nu\alpha\beta}
\, x^{\alpha}x^{\beta}$, where $C_{\mu\nu\alpha\beta}$ is a certain constant tensor
such that $C_{\mu\nu\alpha\beta}=C_{[\mu\nu](\alpha\beta)}$. It is easy to compute that
$Q^{\scriptscriptstyle(2)}{}_{\mu\nu,\sigma}=C_{\mu\nu\sigma\beta}x^{\beta}$,
and it follows that $\xi^{\scriptscriptstyle(2)}{}_{\nu} =
\eta^{\mu\alpha}C_{\mu\nu\alpha\beta}\, x^{\beta}=k_{\nu\beta}\, x^{\beta}$.
Due to the freedom of choice of $x^{\beta}$ we can write
$\eta^{\mu\alpha}C_{\mu\nu\alpha\beta}
=k_{\nu\beta}$. Let us try to decompose $C_{\mu\nu\alpha\beta}$
in a similar way as before: as
the sum of "trace" and "traceless" parts. Let
\begin{equation}\label{Mink1}
    C_{\mu\nu\alpha\beta}= a\eta_{\alpha[\mu}k_{\nu]\beta}
    + a\eta_{\beta[\mu}k_{\nu]\alpha} + bk_{\mu\nu}\eta_{\alpha\beta}
    + D_{\mu\nu\alpha\beta}\ ,
\end{equation}
where $a$ and $b$ are constants, which we may choose freely,
and $D_{\mu\nu\alpha\beta}$ is a certain constant tensor. The above formula and
the properties of
$C_{\mu\nu\alpha\beta}$ imply that $D_{\mu\nu\alpha\beta}=
D_{[\mu\nu](\alpha\beta)}$. We  also have
\[
k_{\nu\beta}=\eta^{\mu\alpha}
C_{\mu\nu\alpha\beta}=(2a-b)k_{\nu\beta} +
\eta^{\mu\alpha}D_{\mu\nu\alpha\beta}\ .
\]
Choosing $a$ and $b$ such that $2a-b=1$, we get
$\eta^{\mu\alpha}D_{\mu\nu\alpha\beta}=0$.
If we use the formula $Q^{\scriptscriptstyle(2)}
{}_{\mu\nu}= \frac12C_{\mu\nu\alpha\beta}\, x^{\alpha}x^{\beta}$
in equation~(\ref{CYK_eq2}) then (after using the decomposition~(\ref{Mink1})
and ordering of terms) we obtain:
\[
D_{\mu\nu\alpha\beta} + D_{\alpha\nu\mu\beta} = \left(\frac23-a\right)
\left(\eta_{\mu\alpha}
k_{\nu\beta} - \frac12\eta_{\nu\alpha}k_{\mu\beta} - \frac12\eta_{\mu\nu}
k_{\alpha\beta} - \frac32\eta_{\beta\mu}k_{\nu\alpha}
- \frac32\eta_{\alpha\beta}k_{\nu\mu}\right) \, .
\]
It is easily seen that if we assume $a=\frac23$ (which means $b=2a-1=\frac13$),
then $D_{\mu\nu\alpha\beta}$ has to satisfy the equation
$D_{\mu\nu\alpha\beta} = - D_{\alpha\nu\mu\beta}$.
Taking into account that $D_{\mu\nu\alpha\beta}=D_{[\mu\nu](\alpha\beta)}$, we get:
\[
D_{\mu\nu\alpha\beta}=-D_{\alpha\nu\mu\beta}=D_{\nu\alpha\beta\mu}=
-D_{\beta\alpha\nu\mu}=D_{\alpha\beta\mu\nu}\ .
\]
But that means that $D_{\mu\nu\alpha\beta}$ is at the same time symmetric
and skew-symmetric in indices $\alpha$ and $\beta$
(and in $\mu$ and $\nu$ as well),
and therefore is identically equal to zero. Finally, we get that
{\setlength\arraycolsep{2pt}
\begin{eqnarray}
Q^{\scriptscriptstyle(2)}{}_{\mu\nu}{} & = &
\left(
\frac13\eta_{\alpha[\mu}k_{\nu]\beta} + \frac13\eta_{\beta[\mu}k_{\nu]\alpha}
+ \frac16k_{\mu\nu}\eta_{\alpha\beta}\right)x^{\alpha}x^{\beta}\nonumber\\
& = & \frac23x_{[\mu} k_{\nu]\beta}x^{\beta} + \frac16k_{\mu\nu}x_{\beta}x^{\beta} \ .
\nonumber
\end{eqnarray}}

Summarizing the above considerations, we can write the general form of a solution of
 equation~(\ref{CYK_eq2}) in Minkowski spacetime:
\begin{equation}\label{CYK_Mink}
    Q^{\mu\nu} = q^{\mu\nu} + \frac23x^{[ \mu}u^{\nu ]}
    +\frac13\varepsilon^{\mu\nu}{}_{\kappa\lambda} v^{\kappa}x^{\lambda}
    +\frac23x^{[\mu} k^{\nu]\lambda}x_{\lambda}
    + \frac16k^{\mu\nu}x_{\lambda}x^{\lambda} \, ,
\end{equation}
where $q^{\mu\nu}$ and $k^{\mu\nu}$ are any constant
skew-symmetric tensors, and $u^{\mu}$ and~$v^{\mu}$ are any
constant vectors. $\xi^{\nu} =Q^{\mu\nu}{}_{,\mu}$ for
$Q^{\mu\nu}$ given by formula~(\ref{CYK_Mink}) is equal to
$\xi^{\mu}=u^{\mu} + k^{\mu\nu}x_{\nu}$. Obviously each
tensor $q^{\mu\nu}$ and $k^{\mu\nu}$ has six independent
components, and each vector $u^{\mu}$ and~$v^{\mu}$ has four
independent components, which means that the space of solutions of
 equation~(\ref{CYK_eq2}) in Minkowski spacetime is
twenty-dimensional\footnote{This is the maximal number of solutions
which is achieved by any conformally flat manifold. In
general, for the CYK $p$--forms in an $n$--dimensional Riemannian manifold $M$
 this number equals $\left(\begin{array}{c} n+2\\ p+1
\end{array} \right)$. Moreover, if a manifold $M$ admits the maximal
possible number of linear independent conformal Yano--Killing
$p$-forms, then it is conformally flat.}.

Let us denote by ${\cal D}$ a {\em dilation vector field}:
\begin{equation}\label{dil_generators}
{\cal D} := x^{\mu}\frac{\partial}{\partial x^{\mu}}\, ,
\end{equation}
which is involved together with Killing fields ${\cal T}_\mu$ and~${\cal
L}_{\mu\nu}$ (cf. (\ref{kvf_generators})) in
the following commutation relations:
\[
[{\cal T}_\mu, {\cal D}] = {\cal T}_\mu, \quad [{\cal D},{\cal L}_{\alpha\beta}] = 0,
\]
\[
[{\cal T}_\mu, {\cal T}_\nu] = 0, \quad \ [{\cal T}_\mu,{\cal L}_{\alpha\beta}]=
\eta_{\mu\alpha}{\cal T}_\beta - \eta_{\mu\beta}{\cal T}_\alpha,
\]
\[
[{\cal L}_{\mu\nu},{\cal L}_{\alpha\beta}] = \eta_{\mu\alpha}{\cal L}_{\beta\nu}
-\eta_{\mu\beta}{\cal L}_{\alpha\nu} + \eta_{\nu\alpha}{\cal L}_{\mu\beta}
-\eta_{\nu\beta}{\cal L}_{\mu\alpha} .
\]
Obviously ${\cal T}_\mu$ and ${\cal L}_{\mu\nu}$ are generators of
the Poincar\'e group.
After adding ${\cal D}$, we get a set of generators of the
pseudo-similarity group (Poincar\'e group extended by scaling transformation
generated by ${\cal D}$).
Formula~(\ref{CYK_Mink}) implies that the following tensors (for $\mu<\nu$)
form a basis in the space of solutions of  equation~(\ref{CYK_eq4}):
\begin{equation}\label{CYK_Mink_base}
{\cal T}_\mu\wedge{\cal T}_\nu\:,\quad {\cal D}\wedge{\cal T}_\mu\:,
\quad \ast({\cal D}\wedge{\cal T}_\mu) \, ,
\quad  {\cal D}\wedge{\cal L}_{\mu\nu}-\frac 12 \eta({\cal D},{\cal D})
    {\cal T}_\mu\wedge{\cal T}_\nu \, .
\end{equation}
Ten of the above CYK tensors, namely ${\cal T}_\mu\wedge{\cal T}_\nu$
and~$\ast({\cal D}\wedge{\cal T}_\mu)$, are Yano tensors.

Using the basis~(\ref{CYK_Mink_base}) we can represent any CYK tensor
in Minkowski spacetime as follows:
\begin{equation}
Q  =  \tilde{q}^{\mu\nu}{\cal T}_\mu\wedge{\cal T}_\nu +
\tilde{u}^{\mu}\,{\cal D}\wedge{\cal T}_\mu\ + \tilde{v}^{\mu}*\!
({\cal D}\wedge{\cal T}_\mu) 
+ \tilde{k}^{\mu\nu}({\cal D}\wedge{\cal L}_{\mu\nu}
- \frac 12 \eta({\cal D},{\cal D}){\cal T}_\mu\wedge{\cal T}_\nu)\, ,\nonumber
\end{equation}
where $\tilde{q}^{\mu\nu}$ and $\tilde{k}^{\mu\nu}$
are certain constant
skew-symmetric tensors, whereas $\tilde{u}^{\mu}$ and~$\tilde{v}^{\mu}$
are certain constant vectors.

Let $W$ be a spin-2 field in Minkowski spacetime. We already know
that every CYK tensor defines a certain conserved quantity related
to the field $W$. Let us write out the charges given by
tensors~(\ref{CYK_Mink_base}). We have:
\begin{equation}\label{charge_w}
    w_{\mu\nu}:= \frac1{16\pi} \int_{\partial\Sigma} W({\cal T}_\mu\wedge{\cal T}_\nu),
\end{equation}

\begin{equation}\label{charge_p}
    p_\mu := \frac1{16\pi}\int_{\partial\Sigma} W({\cal D}\wedge{\cal T}_\mu),
\end{equation}

\begin{equation}\label{charge_b}
    b_\mu :=\frac1{16\pi}\int_{\partial\Sigma} {^*}W({\cal D}\wedge{\cal T}_\mu),
\end{equation}

\begin{equation}\label{charge_j}
    j_{\mu\nu} :=\frac1{16\pi}\int_{\partial\Sigma}W\left({\cal D}\wedge
    {\cal L}_{\mu\nu} -\frac 12 \eta({\cal D},{\cal D})
    {\cal T}_\mu\wedge{\cal T}_\nu \right),
\end{equation}where the notation $W(A\wedge B)$ is to be understood as
$W^{\mu\nu}{}_{\alpha\beta}A^\alpha B^\beta \rd \sigma_{\mu\nu}$.
The factor $\frac1{16\pi}$ was introduced for normalization.

When $W$ is the Weyl tensor of the linearized Einstein theory, $p_\mu$
turns out to be its momentum, $b_\mu$ --- its dual momentum, and
$j_{\mu\nu}$~--- its angular momentum tensor\footnote{These
identifications justify the use of the normalizing
factor~$\frac1{16\pi}$.} (see~\cite{JJspin2}). In general, there
might  also be non-vanishing charges $w_{\mu\nu}$.
However, if we assume that $W =
O(\frac1{r^3})$ (which is a typical situation), then
$w_{\mu\nu}=0$. In this case there are only 14 nontrivial charges but
 we do not know any local (i.e. using only field equation)
argument for this being a general rule. It turns out that
vanishing of $w_{\mu\nu}$ is necessary for defining angular
momentum of the field at future null infinity
(see~\cite{Goldberg1}, \cite{cykem}).

We do not have to restrict ourselves in the above considerations only
to the case of Minkowski spacetime. All that what was written above can
be easily adapted to the case of asymptotically flat spacetime. By
asymptotically flat spacetime we mean a manifold $M$ with a metric
$g$ and a set of coordinates $(x^\mu)$ such that components of $g$
satisfy the following conditions:
\begin{equation}\label{asympt_flat}
g_{\mu\nu} - \eta_{\mu\nu} = O\left(\frac1{r}\right), \quad  \quad
g_{\mu\nu,\kappa} = O\left(\frac1{r^2}\right),
\end{equation}
where $r:=\left(\sum^{3}_{k=1}(x^{k})^2\right)^{\frac12}$, and
$\eta_{\mu\nu}= \textrm{diag}(-1,1,1,1)$. The coordinates $(x^\mu)$ do
not have to  be global coordinates on $M$. It is sufficient for
them to be defined for $r$ greater than some $r_0$ (that means
that each set of four numbers $(x^0, x^1, x^2, x^3)$, for which
$r>r_0$, defines some point of $M$).

Now, let $W$ be a spin-2 field defined on $M$. If the metric
admits solutions of  equation~(\ref{CYK_eq2}), which for $r$
going to infinity go to tensor ${\cal D}\wedge {\cal T}_\mu$, then
we say that~$p_\mu$ defined by the formula~(\ref{charge_p})
defines the $\mu$ component of the momentum of the field~$W$.
Similarly, we can define dual momentum and angular momentum of the
field $W$, provided that the metric $g_{\mu\nu}$ admits CYK
tensors with appropriate asymptotic behavior at spatial infinity.

\section{CYK tensor and related objects in Kerr spacetime}\label{Kerrch}

The Kerr solution to the vacuum Einstein equations is the metric $g$ which
in a certain coordinate system $(t,r,\theta,\phi)$ has the form:
\begin{equation}\label{Kerr_metric}
g = g_{tt} \rd t^2 + 2g_{t\phi} \rd t \rd \phi + g_{rr} \rd r^2
+ g_{\theta\theta} \rd \theta^2 + g_{\phi\phi} \rd \phi^2\, ,
\end{equation}
where
\[
g_{tt} = -1 +{2mr \over \rho ^2},  \quad g_{t\phi} = -{2mra\sin ^2 \theta
\over \rho ^2}, \quad g_{rr} = {\rho ^2 \over \triangle}, \quad
g_{\theta \theta} = \rho^2,
\]
\begin{equation}\label{Kerr_metric_compts}
g_{\phi\phi} =
\sin^2\theta\left(r^2+a^2+{2mra^2\sin^2\theta\over\rho^2}\right),
\end{equation}
with
\begin{equation}\label{Kerr_metric_symbols}
 \rho ^2 = r^2 + a^2 \cos ^2 \theta \quad \textrm{and} \quad
 \triangle = r^2 -2mr +a^2.
\end{equation}
This metric describes a rotating black hole of mass $m$ and angular momentum
\mbox{$J=am$}. The advantage of the above coordinates
(called Boyer--Lindquist coordinates) is that for $r$ much
greater than $m$ and $a$
\[
g \thickapprox -\rd t^2 + \rd r^2 + r^2\rd \theta^2 + r^2\sin^2\theta
\rd\phi^2,
\]
which shows that $g$ approximates the flat metric for $r$ going to infinity.
It is not hard to see that  equations~(\ref{asympt_flat}) are satisfied,
which means that $g$ is an asymptotically flat metric. However, for $r$ equal
to $r_{\pm}:=m\pm \sqrt{m^2-a^2}$ the component $g_{rr}$ is singular
(due to $\Delta(r=r_{\pm})=0$).
A much better coordinate
system for describing the Kerr metric near the surfaces ${r=r_{\pm}}$ is
Eddington--Finkelstein coordinates: $(v,r,\theta,\tilde{\phi})$, where
$r$ and $\theta$ are the same as before whereas $v$ and $\tilde{\phi}$
are related to Boyer--Lindquist coordinates via the following formulae:
\[
\rd v = \rd t + \frac{r^2+a^2}{\Delta}\rd r, \quad
\rd \tilde{\phi} = \rd \phi + \frac{a}{\Delta}\rd r.
\]
In Eddington-Finkelstein coordinates $g$ has the form:
\[
g = \tilde{g}_{vv} \rd v^2 + 2\tilde{g}_{v\tilde{\phi}} \rd v \rd
\tilde{\phi} + 2\tilde{g}_{r\tilde{\phi}}\rd r\rd \tilde{\phi}
+ 2\tilde{g}_{rv} \rd r\rd v + \tilde{g}_{\theta\theta} \rd \theta^2
+ \tilde{g}_{\tilde{\phi}\tilde{\phi}} \rd \tilde{\phi}^2,
\]
where
\[
\tilde{g}_{vv} = g_{tt},\quad \tilde{g}_{v\tilde{\phi}} = g_{t\phi},\quad
\tilde{g}_{r\tilde{\phi}} = -a\sin^2\theta,
\]
\[
\tilde{g}_{rv} = 1, \quad \tilde{g}_{\tilde{\phi}\tilde{\phi}} = g_{\phi\phi},\quad
\tilde{g}_{\theta\theta} = g_{\theta\theta}.
\]
It can be easily seen that now singularities are absent.
Nevertheless, in our considerations we will mostly use
Boyer--Lindquist coordinates.

A natural question arises: what are the solutions of
equation~(\ref{CYK_eq2}) for a rotating black hole described by the Kerr
metric? Finding the form of the general CYK tensor in Minkowski
spacetime turned out to be quite an easy task, but now the situation is
not so simple, as we are dealing with quite a complicated
overdetermined system of differential equations for components
$Q_{\mu\nu}$. However, there is known in the literature
(see~\cite{Gib-Riet-vHolt}, \cite{PenKY}, \cite{Pen-Walk}) one
solution\footnote{According to \cite{HS} this is also valid
for the charged Kerr solution. Penrose and Walker suggest in \cite{Pen-Walk}
than it may be even generalized to the case with cosmological constant.
For example, it would be nice to check explicitly existence of CYK tensor
for Kerr-Newman-AdS black hole.}
\begin{equation}\label{Kerr_Q}
Y:=Q_{{\rm Kerr}}=
r\sin\theta \rd\theta \wedge \left[ \left( r^2+a^2\right)\rd\phi - a\rd t
\right]+ a\cos\theta \rd r\wedge\left(\rd t - a\sin^2\theta \rd\phi \right).
\end{equation}
Strictly speaking, it is a solution\footnote{To avoid confusion we denote
this particular solution by $Y$.} of  equation~(\ref{Yano_eq})
(i.e. Yano tensor), which -- as we know --
is a special case of  equation~(\ref{CYK_eq2}). It means that
$\xi^{\mu}:=Y^{\nu\mu}{}_{;\nu}=0$.

Since the Kerr metric is asymptotically flat, $Y_{\mu\nu}$ in the
neighborhood of spatial infinity should be of the
form~(\ref{CYK_Mink}). Indeed, asymptotically this tensor looks as
follows:
\begin{displaymath}
Y = r^3\sin\theta \rd\theta \wedge \rd\phi + O\left(1\right) =
\ast({\cal T}_{0} \wedge {\cal D}) + O\left(1\right).
\end{displaymath}

From (\ref{charge_b}) we see that a charge related to $Y_{\mu\nu}$
is simply $-b_0=b^0$, which represents dual energy. The Riemann tensor
$R_{\mu\nu\rho\sigma}$ of (Ricci flat) Kerr spacetime is
simultaneously a Weyl tensor and fulfils
properties from Definition~\ref{spin2_df} of a spin-2 field. Hence we obtain:
{\setlength\arraycolsep{2pt}
\begin{eqnarray}\label{Kerr_F}
F(R,Y) & = & \frac{4ma\sin\theta\left(r^2-a^2\cos^2\theta\right)}
{\rho ^4} \rd t\wedge \rd\theta + \frac{8mar\cos\theta} {\rho
^4}\rd t\wedge \rd r
\nonumber\\
& & {}+
\frac{4m\sin\theta\left(r^2+a^2\right)\left(r^2-a^2\cos^2\theta\right)}
{\rho ^4}\rd\theta \wedge \rd\phi \nonumber\\
& & {} + \frac{8ma^2r\cos\theta\sin^2\theta}{\rho ^4}\rd r \wedge
\rd \phi \, ,
\end{eqnarray}}\noindent
where according to~(\ref{def_FWQ}) $F_{\mu\nu}(R,Y) :=
R_{\mu\nu\lambda\kappa}Y^{\lambda\kappa}$. We have used symbolic
software like {\sc Mathematica} and {\sc Waterloo Maple} to check
the result \eq{Kerr_F}.

Direct integration of $F(R,Y)$ on a sphere $r=\textrm{const.}$
and~$t=\textrm{const.}$ gives
\[
\frac{1}{16\pi}\int_{S} F^{\mu\nu}(R,Y)\rd \sigma_{\mu\nu} = 0\, ,
\]
which means that the dual energy of the Kerr black hole vanishes.

Theorem~\ref{dual_th} implies that the dual tensor:
\begin{equation}\label{Kerr_Qstar}
    \ast Y=a\cos\theta\sin\theta \rd\theta \wedge \left[ \left(r^2+a^2
    \right) \rd\phi - a\rd t \right]+ r\rd r \wedge \left(a\sin^2\theta \rd\phi -
    \rd t \right)
\end{equation}
is also a CYK tensor in Kerr spacetime.
The two-form $\ast Y$ is closed
\footnote{$Y$
is a Killing-Yano tensor and corresponding covector $\xi$ vanishes.}
(i.e. $\rd \ast Y=0$). On the other hand,
$ \rd Y= 3\rho^2\sin\theta\rd\theta\wedge\rd\phi\wedge\rd r=
     3\partial_{t} \lrcorner \Omega=3\rd\Sigma_{t} $
(where $\Omega=\rho^2\sin\theta\rd t\wedge\rd\theta\wedge\rd\phi\wedge\rd r$ is
a volume form)
     and a vector field
$\chi=\ast Y^{\mu\nu}{}_{;\mu}\partial_\nu=3\partial_{t}$ is not
vanishing.
This implies that $\ast Y$ is a
{\em Killing potential} which satisfies the CYK equation (cf. \cite{Glass-Kress}).

The asymptotics of tensor $*Y$ look as follows:
\begin{eqnarray*}
\ast Y = r\partial_{r}\wedge\partial_{t} +
O\left(\frac{1}{r}\right) = {\cal D}\wedge {\cal T}_{0} +
O\left(\frac{1}{r}\right)\, .
\end{eqnarray*}
From~(\ref{charge_p}) we obtain that the charge related to $*Y$ is $p_0$
and corresponds to the energy.
We have: {\setlength\arraycolsep{2pt}
\begin{eqnarray}\label{Kerr_Fstar}
F(R,\ast Y) & = & \frac{8ma^2r\sin\theta\cos\theta} {\rho ^4}\rd
\theta \wedge \rd t + \frac{4m(r^2-a^2\cos^2\theta)} {\rho ^4}\rd
t\wedge \rd r  \nonumber\\
& & {}+ \frac{8mar\sin\theta\cos\theta(r^2+a^2)} {\rho
^4}\rd\phi \wedge \rd\theta \nonumber\\
& & {} + \frac{4ma\sin^2\theta(r^2-a^2\cos^2\theta)}{\rho ^4}\rd r
\wedge \rd \phi \, .
\end{eqnarray}}
Integration over a sphere $r=\textrm{const.}$ and~\mbox{$t=\textrm{const.}$}
gives
\be\label{mKerr}
\frac{1}{16\pi}\int_{S} \ast F^{\mu\nu}(R,Y)\rd
\sigma_{\mu\nu} = -m \, ,
\ee
which means that the total energy (mass) of the Kerr black hole equals $m$.
Formula \eq{mKerr} expresses mass of a rotating black hole quasi-locally.

Let us denote by $A$ and $A'$ the potentials of the
closed two-forms $F(R,Y)$ and $F(R,*Y)$  respectively (cf. \ref{Maxwell_eq2})),
i.e. $F(R,Y) = \rd A$ and $F(R,*Y)= \rd A'$.
 In Boyer--Lindquist coordinates they look as follows
\begin{equation}\label{Kerr_A}
A = \frac{4ma\cos\theta}{\rho^2}\rd t -
\frac{4m\cos\theta\left(r^2+a^2\right)} {\rho^2}\rd\phi,
\end{equation}
\begin{equation}\label{Kerr_A'}
A' = \left(\frac{4mr}{\rho^2} - 2\right)\rd t -
\frac{4mar\sin^2\theta}{\rho^2}\rd\phi.
\end{equation}
The one-forms $A$ and $A'$ are always well defined locally.
It turns out that potential
$A'$ is global, but $A$ is well defined except at $\theta=0$ and
$\theta=\pi$. Formulae (\ref{Kerr_A})--(\ref{Kerr_A'}) are obviously
not unique, the potentials $A$ and $A'$ are given
up to the gauge $A \to A+\rd f$ and $A' \to A'+\rd f'$, where $f$ and $f'$ are
arbitrary functions. We would like to stress that $A$ is
essentially singular, i.e. all potentials for
 $F(R,Y)$ have the same singularity\footnote{Let us denote by $B$ a gauge
 equivalent potential to $A$ i.e. $B=A+\rd f$.
 Integrating $A$ and~$B$ along a curve $t=\textrm{const.}$,
$r=\textrm{const.}$, $\theta=\textrm{const.}$, $0<\phi<2\pi$
we obtain the same result because the function $f$ has to be
periodic, $f(\phi=2\pi)-f(\phi=0)=0$. For regular $B$
in the limit $\theta\rightarrow 0$
or $\theta\rightarrow \pi$ our curve goes to a point and the corresponding
integral has to vanish. On the other hand, integrating $A$
we obtain $8\pi
m\cos\theta\left(r^2+a^2\right)/\rho^2$, which converges to $8\pi
m\left(r^2+a^2\right) /\rho^2$ for $\theta\to 0$ or $\theta\to\pi$.}.
The result is not surprising because the two-form
 $-F(R,*Y)$ describes the ``electric'' charge $m$ hidden under horizon.
Its dual two-form $F(R,Y)$ describes the ``magnetic'' charge and corresponds
to a Dirac monopole. This explains why
  potential $A$ is not global.

Passing from covectors $A$ and $A'$ to vector fields we obtain
\begin{equation}\label{Kerr_Avect}
g^{-1}(A) = -\frac{4ma\cos\theta}{\rho^2}\partial_{t} -
\frac{4m\cos\theta} {\rho^2\sin^2\!\theta}\partial_{\phi}
\end{equation}
and
\begin{equation}\label{Kerr_A'vect}
g^{-1}(A') = 2\partial_{t}.
\end{equation}
In particular, $g^{-1}(A')$ is a Killing vector field which implies that
 $F(R,*Y)$ is a Papapetrou field\footnote{Papapetrou~(cf.~\cite{Fay-Sop},
 \cite{Papa}) pointed out that if $K$ is a
 Killing vector field for metric~$g$ then
$F:=\rd\left(g(K)\right)$ fulfils the Maxwell equations with current
$j^\mu:=R^\mu{}_\nu K^\nu$, where $R_{\mu\nu}$ is
Ricci tensor. In particular, for Ricci flat metric  $g$
 the two-form $F$ is a solution to  vacuum Maxwell equations.}.

The Kerr metric is a special case of the electrovac Kerr--Newman metric\footnote{In
 Boyer--Lindquist coordinates we have
\[
g_{\textrm{\tiny K-N}} = -\frac{\Delta}{\rho^2}\left(\rd t - a\sin^2\theta\rd
\phi\right)^2 + \frac{\sin^2\theta}{\rho^2}\left((r^2+a^2)\rd \phi
- a\rd t\right)^2 + \frac{\rho^2}{\Delta}\rd r^2 + \rho^2 \rd
\theta^2,
\]
where $\Delta = r^2 - 2mr + a^2 + q^2$ and $\rho^2 = r^2 + a^2
\cos^2\theta$. Moreover, the
vector potential $\displaystyle A_{\textrm{\tiny K-N}}= qr
\left(\rd t - a\sin^2\theta\rd\phi\right)/{\rho^2}$
 is associated with the electromagnetic two-form
$F_{\textrm{\tiny K-N}}=\rd A_{\textrm{\tiny K-N}}$.
Obviously, for $q=0$ we obtain the Kerr metric.}
describing a rotating black hole with electric charge
 $q$. It turns out that
the two-form $F(R,*Y)$ (given by~(\ref{Kerr_Fstar})) is related
to the Maxwell field $F_{\textrm{\tiny K-N}}$
 describing electromagnetic field in Kerr--Newman spacetime.
 More precisely, both fields have the same form
 up to the exchange of constants $q$ and $-4m$.
For the Kerr--Newman solution the corresponding electromagnetic tensor
$F_{\textrm{\tiny K-N}}$ has the following property:
\begin{displaymath}
F(W,\ast Y) = -\lim_{q\rightarrow 0}\frac{4m}{q} F_{\textrm{\tiny K-N}}\, .
\end{displaymath}

Finishing this Section let us pass to Eddington--Finkelstein
coordinates. Formulae~(\ref{Kerr_Q})--(\ref{Kerr_Fstar})
and~(\ref{Kerr_Avect})--(\ref{Kerr_A'vect}) take a similar form
which corresponds to the change of the names of
coordinates from  $t$ to $v$ and $\phi$ to $\tilde{\phi}$ respectively.
We have
\[
Y=r\sin\theta \rd\theta \wedge \left[ \left(
r^2+a^2\right)\rd\tilde{\phi} - a\rd v \right]+ a\cos\theta \rd
r\wedge\left(\rd v - a\sin^2\theta \rd\tilde{\phi} \right),
\]
\[
    \ast Y=a\cos\theta\sin\theta \rd\theta \wedge \left[ \left(r^2+a^2
    \right) \rd\tilde{\phi} - a\rd v \right]+ r\rd r \wedge \left(a\sin^2\theta
    \rd\tilde{\phi} - \rd v \right),
\]
etc. Let us notice that the same procedure applied
to equations~(\ref{Kerr_A})--(\ref{Kerr_A'}) leads to
\[
{\bar A} = \frac{4ma\cos\theta}{\rho^2}\rd v -
\frac{4m\cos\theta\left(r^2+a^2\right)} {\rho^2}\rd\tilde{\phi}
\]
and
\[
{\bar A}' = \left(\frac{4mr}{\rho^2} - 2\right)\rd v -
\frac{4mar\sin^2\theta}{\rho^2}\rd\tilde{\phi} \, ,
\]
which are not equal to $A$ and $A'$ respectively.
However, they are gauge equivalent and describe respectively the same fields
 $F(R,Y)$ and $F(R,*Y)$.

We denote the conformal Killing tensors associated with $Y$ and $*Y$ by
$K_{\mu\nu}:=Y_{\mu}{}^{\kappa}Y_{\kappa\nu}$,
$K'_{\mu\nu}:=*Y_{\mu}{}^{\kappa} *\!Y_{\kappa\nu}$
and $K''_{\mu\nu}:= \frac12 \left( Y_{\mu}{}^{\kappa} *\!Y_{\kappa\nu} +
*Y_{\mu}{}^{\kappa} Y_{\kappa\nu} \right)$.
Theorem~\ref{killing_tens_th} implies that $K'$ and $K''$ are
conformal Killing tensors and, moreover, $K$ is a
Killing tensor.
 In Eddington--Finkelstein coordinates (with raised indices)
 they look as follows: 
  {\setlength\arraycolsep{2pt}
\begin{eqnarray} \label{KTK}
K & = & -\frac{a^2r^2\sin^2\theta}{\rho^2}\partial_v \!\otimes \partial_v -
\frac{r^2}{\rho^2}\partial_\theta \!\otimes \partial_\theta -
\frac{r^2}{\rho^2\sin^2\theta}\partial_{\scriptscriptstyle\tilde{\phi}}
\!\otimes \partial_{\scriptscriptstyle\tilde{\phi}}
+ \frac{a^2 \Delta \cos^2\theta}{\rho^2} \partial_r \!\otimes \partial_r
\nonumber \\[10pt]
& & - \frac{2ar^2}{\rho^2}\partial_v \!\otimes_{s}\!
\partial_{\scriptscriptstyle\tilde{\phi}} +
\frac{2a^2(r^2+a^2)\cos^2\theta}{\rho^2}\partial_v \!\otimes_{s}\!  \partial_r +
\frac{2a^3\cos^2\theta}{\rho^2}\partial_r \!\otimes_{s}\!
\partial_{\scriptscriptstyle\tilde{\phi}}\,, 
\end{eqnarray} }
where $a\!\otimes_s b:= \frac12 (a\!\otimes b + b\!\otimes a)$,

{\setlength\arraycolsep{2pt}
\begin{eqnarray}
K' & = & -\frac{a^4\sin^2\theta\cos^2\theta}{\rho^2}\partial_v\!\otimes\partial_v
-\frac{a^2\cos^2\theta} {\rho^2}\partial_\theta\!\otimes\partial_\theta -
\frac{a^2\cos^2\theta}{\rho^2\sin^2\theta}
\partial_{\scriptscriptstyle\tilde{\phi}}\!\otimes
\partial_{\scriptscriptstyle\tilde{\phi}} +
\frac{r^2\Delta}{\rho^2}\partial_r\!\otimes\partial_r
\nonumber \\[10pt]
& & - \frac{2a^3\cos^2\theta}{\rho^2}\partial_v \!\otimes_{s}\!
\partial_{\scriptscriptstyle\tilde{\phi}} +
\frac{2r^2(r^2+a^2)}{\rho^2}\partial_v \!\otimes_{s}\! \partial_r
+ \frac{2ar^2}{\rho^2}\partial_r \!\otimes_{s}\!
\partial_{\scriptscriptstyle\tilde{\phi}}  
\end{eqnarray} }
and 
{\setlength\arraycolsep{2pt}
\begin{eqnarray}
K'' & = & -\frac{a r \cos\theta}{\rho^2} \Big(
a^2\sin^2\theta \, \partial_v\!\otimes\partial_v +
\partial_\theta\!\otimes\partial_\theta +
\frac{1}{\sin^2\theta}
\partial_{\scriptscriptstyle\tilde{\phi}}\!\otimes
\partial_{\scriptscriptstyle\tilde{\phi}} + \Delta
\partial_r\!\otimes\partial_r
\nonumber \\[10pt]
& & + 2a \partial_v \!\otimes_{s}\!
\partial_{\scriptscriptstyle\tilde{\phi}} +
2(r^2+a^2)\partial_v \!\otimes_{s}\! \partial_r
+ 2a \partial_r \!\otimes_{s}\!
\partial_{\scriptscriptstyle\tilde{\phi}} \Big)\,. 
\end{eqnarray} }
In particular, the traces of our conformal Killing tensors are the following:
\[
 - \tr K = \tr K' = 2\left(r^2 - a^2\cos^2\theta\right) \, ,
\quad
\tr K'' = -4ar\cos\theta \, .
\]
The Killing tensor $K^{\mu\nu}$
 given in \eq{KTK} is the one  which was found in \cite{Pen-Walk}
 (see also \cite{Casalbuoni},  \cite{Gib-Riet-vHolt},
 \cite{PenKY}).
The constant of motion $K^{\mu\nu}p_\mu p_\nu$
along geodesics, given by
\be\label{rg} g^{\mu\nu} p_\mu \displaystyle\nabla_{\nu} \, p = 0 \, , \ee
is called the Carter constant and
it is indispensable for the full description of geodesics in Kerr spacetime
 (cf. \cite{Carter}, \cite{deFelice}, \cite{Frol-Nov}, \cite{Pen-Walk}).
In this context, our two conformal Killing tensors\footnote{Unlike
 the Killing tensor $K$ they seem not to be
available in the literature in explicit form.} $K'$ and $K''$
 produce similar constants of motion along null geodesics.
 More precisely, \eq{CK_tens_eq} together with the geodesic
 equation \eq{rg} implies that for a null covector $p$ and for any conformal
 Killing tensor $K$ we have
 \[ g^{\mu\nu} p_\mu \displaystyle\nabla_{\nu}
 (K^{\mu\nu}p_\mu p_\nu) =0 \, ,\]
 i.e. $A^{\mu\nu}p_\mu p_\nu$ is a constant of motion along null geodesics.

\section{Conserved quantities bilinear in terms of spin-2 field}

Let us start with the standard definition of the energy-momentum
tensor for a Maxwell field $F$:
\be\label{TEM}
 T^{\rm\scriptscriptstyle EM}_{\mu\nu}(F) :=
 \frac1{8\pi} \left( F_{\mu\sigma}F_\nu{^\sigma}
  + F^*{_{\mu\sigma}}F^*{_\nu{^\sigma}}
 \right) = \frac1{4\pi}\left(F_{\mu\sigma}F_\nu{^\sigma} -\frac14 g_{\mu\nu}
 F_{\sigma\rho}F^{\sigma\rho}\right) \, .
\ee
The energy-momentum tensor $T^{\rm\scriptscriptstyle
EM}_{\mu\nu}(F)$ is symmetric, traceless and satisfies the
following positivity condition: {\em for any non-spacelike
future-directed vector fields $X,Y$ we have
$T^{\rm\scriptscriptstyle EM}_{\mu\nu}(F)X^\mu Y^\nu \geq 0\, .$}
Straightforwardly from the definition we get
\be\label{TF*F}
 T^{\rm\scriptscriptstyle EM}_{\mu\nu}(F) =
 T^{\rm\scriptscriptstyle EM}_{\mu\nu}(F^*) \, . \ee
Moreover, if $F$ is a Maxwell field then
\be\label{divTEM}
\nabla^\mu T^{\rm\scriptscriptstyle EM}_{\mu\nu}(F)=0 \, ,
 \ee
and if $X$ is a conformal Killing vector field then the quantity
\[ \QEM(X,\Sigma;F):=
\int_\Sigma T^{\rm\scriptscriptstyle
EM}_{\mu\nu}X^\mu\rd\Sigma^\nu \] defines a global conserved
quantity
 for the spacelike hypersurface $\Sigma$ with the
end at spacelike infinity.

Let us restrict ourselves to the spacelike hyperplanes $\Sigma_t
:= \{ x\in M \; : \; x^0=t=\mbox{const.} \}$. We use the following
convention for indices: $(x^\mu)$ $\mu=0,\dots,3$ are Cartesian
coordinates in Minkowski spacetime, $x^0$ denotes a temporal
coordinate and $(x^k)$ $k=1,2,3$ are coordinates on the spacelike
surface $\Sigma_t$. If the quantity $\QEM(X,\Sigma_t)$ is finite
for $t=0$ than it is constant in time. If we want to get a
positive definite integral $\QEM$, we have to restrict ourselves
to the case of a non-spacelike field $X$. We can choose
 time translation ${\cal T}_0$ or
time-like conformal acceleration ${\cal K}_0$, where
\be\label{Kcf}
 {\cal K}_\mu := -2x_\mu{\cal D} + x^\sigma x_\sigma {\cal T}_\mu \ee
is a set of four ``pure'' conformal Killing vector fields which
should be added to the eleven fields (\ref{kvf_generators}) and
(\ref{dil_generators}) to obtain the full
15-dimensional algebra of the conformal group. This way we get
\begin{Theorem}\label{HT+HK}
 There exist two conserved (time-independent) positive definite integrals
 $\QEM( {\cal T}_0,\Sigma_t; F)$ and $\QEM( {\cal K}_0,\Sigma_t; F)$
 for a field $F$ satisfying the vacuum Maxwell equations.
\end{Theorem}
\begin{proof}
This is a simple consequence of \eq{divTEM} and the traceless property
of $\TEM$ which implies $\nabla^\mu \left(
T^{\rm\scriptscriptstyle EM}_{\mu\nu}X^\nu\right)=0$ for any
conformal Killing vector field $X$.
\end{proof}

Following \cite{Ch-Kl}, for the Bel--Robinson tensor defined as
follows:
 \ber\label{BRt} T^{\scriptscriptstyle
BR}_{\mu\nu\kappa\lambda} & := & W_{\mu\rho\kappa\sigma}
W_\nu{^\rho}{_\lambda}{^\sigma} +
{W^*}_{\mu\rho\kappa\sigma} {W^*}_\nu{^\rho}{_\lambda}{^\sigma} \\
& =&
 W_{\mu\rho\kappa\sigma} W_\nu{^\rho}{_\lambda}{^\sigma}
 + W_{\mu\rho\lambda\sigma} W_\nu{^\rho}{_\kappa}{^\sigma}
-\frac18 g_{\mu\nu}g_{\kappa\lambda}
W_{\alpha\beta\gamma\delta}W^{\alpha\beta\gamma\delta} \, , \eer
where $W$ is a spin-2 field, one can find a natural generalization
of Theorem \ref{HT+HK} which is a consequence of the
properties similar to (\ref{divTEM}). More precisely,
$T^{\scriptscriptstyle BR}_{\mu\nu\kappa\lambda}$ is symmetric and
traceless in all pairs of indices. Moreover,
\[ T^{\scriptscriptstyle BR}(W)=T^{\scriptscriptstyle BR}(W^*) \, , \]
and if $W$ is a spin-2 field than \be\label{divTBR} \nabla^\mu
T^{\rm\scriptscriptstyle BR}_{\mu\nu\lambda\kappa}(W)=0 \, .\ee If
$X,Y,Z$ are conformal Killing vector fields then the quantity
\[ \QBR(X,Y,Z,\Sigma_t;W):=
\int_{\Sigma_t} T^{\rm\scriptscriptstyle
BR}_{\mu\nu\lambda\kappa}X^\mu Y^\nu Z^\lambda \rd\Sigma^\kappa \]
defines a global charge at time $t$. The quantity
$T^{\scriptscriptstyle BR}(X,Y,Z,T)$ is non-negative for any
non-spacelike future-directed vector fields $X,Y,Z,T$ whenever at
most two of the vector fields are distinct. From the above properties
we obtain an extension
 of  Theorem \ref{HT+HK} to the case of a spin-2 field $W$:
\begin{Theorem}\label{HTTT+HKTT}
There exist four conserved (time-independent) positive definite integrals
 $\QBR({\cal T}_0,{\cal T}_0, {\cal T}_0,\Sigma_t; W)$,
 $\QBR({\cal K}_0,{\cal T}_0, {\cal T}_0,\Sigma_t; W)$,
 $\QBR({\cal K}_0,{\cal K}_0, {\cal T}_0,\Sigma_t; W)$  and \\
 $\QBR({\cal K}_0,{\cal K}_0, {\cal K}_0,\Sigma_t; W)$
 for the spin-2 field $W$ satisfying field equations \eq{Bianchi_id}.
\end{Theorem}
\begin{proof}
Similarly as in Theorem \ref{HT+HK}, from \eq{divTBR} and the
traceless property of $\TBR$ we get $\nabla^\mu
\left(T^{\rm\scriptscriptstyle BR}_{\mu\nu\lambda\kappa}X^\kappa
Y^\nu Z^\lambda \right)=0$ for any conformal Killing vector fields
$X,Y,Z$.
\end{proof}

In \cite{cykem} a generalization of the above
considerations led to the following functional:
 \be\label{CQYK}
  \QYK(X,V;Q):= \int_{{V\subset\Sigma}}
   T^{\rm\scriptscriptstyle EM}_{\mu\nu}
   \bigl(F(W,Q)\bigr)X^\mu\rd\Sigma^\nu
 \ee
 and according to Theorem 3 in \cite{cykem},
 {\em the four conserved quantities $\QBR$ 
 are contained in the functionals $\QYK$.}\\
 In particular, the functional
\be \label{EMV}
\QYK_0  :=  \int_V \TEM\bigl({\cal T}_0,{\cal T}_0, F(W,{\cal
D}\wedge{\cal T}_0)\bigr)\rd^3 x
 \ee
 has a positive integrand which is a natural
 candidate for a certain ``energy density''.
It contains only first (radial and time) derivatives
of quasi-local variables describing linearized gravity.
 The functional $\QYK_0$
is also very close to the reduced Hamiltonian proposed in
\cite{JJschwarzl}. They are proportional to each other
after spherical harmonic decomposition
 (i.e. for each spherical mode).

\section{Density of ``spin energy'' for rotating black hole}

The ADM mass of the Kerr black hole \eq{Kerr_metric_compts}
given by the quasi-local formula \eq{mKerr} equals $m$.
On the other hand, the {\it irreducible mass} of the horizon
(see \cite{mirr}, \cite{Frol-Nov})
is defined by the quadratic relation to its two-dimensional area $A_H$:
\[ M_{irr}=\sqrt{\frac{A_H}{16\pi}} =
\frac12\sqrt{r_{+}^2+a^2}=\frac12\sqrt{2mr_{+}}\, . \]
The energy $M_{irr}$ plays the role
of a lower bound in Penrose inequality (cf. \cite{PenIneq}, \cite{procesRP})
and never decreases according to the second law of black hole physics
(see e.g. \cite{Frol-Nov} or \cite{Heusler}).
The difference between total ADM mass $m$ and the irreducible mass $M_{irr}$
is often interpreted as {\it rotational energy}
\[ M_{rot} := m-M_{irr}= m \left( 1-\sqrt{r_+\over 2m} \right) \approx a^2/8m
\, . \]
Contrary to $M_{irr}$ the energy $M_{rot}$ is accessible by
reversible transformations like the Penrose process
\cite{procesRP} which takes place in the (ergo-)region\footnote{Usually
called \emph{ergosphere} which is completely misleading
because it is \underline{not} a sphere.} between
event horizon and ergo-surface\footnote{It is given by the largest root
of $\Delta(r)=a^2\sin^2\theta$, usually called the \emph{stationary limit surface}.
Moreover, the instantaneous ergo-surface is a two-dimensional manifold with
spherical topology and conical singularity on the axis $(\theta=0,\pi)$
(cf. \cite{lake}).}.

The two-form \eq{Kerr_Fstar}, previously used in Section \ref{Kerrch}
to obtain the quasi-local mass, may be also
  applied to the construction of a quadratic functional \eq{EMV}
  denoted by $\QYK_0$.
Let us consider an auxiliary Maxwell field $f$ which is a solution
of the vacuum Maxwell equations on the Kerr background and is given by
{\setlength\arraycolsep{2pt}
\begin{eqnarray}\label{Kerrf}
 f_{aux}=F(R,\ast Y)/4m & = &
\frac{2a^2r\sin\theta\cos\theta}{\rho ^4}\rd \theta \wedge \rd t +
\frac{r^2-a^2\cos^2\theta}{\rho ^4}\rd
t\wedge \rd r  \nonumber\\
& & {}+ \frac{2ar\sin\theta\cos\theta(r^2+a^2)} {\rho^4}
\rd\phi \wedge \rd\theta \nonumber\\
& & {} + \frac{a\sin^2\theta(r^2-a^2\cos^2\theta)}{\rho ^4}\rd r
\wedge \rd \phi \, .
\end{eqnarray}}

The density of the energy built from $f$ and corresponding to
the integrand in the functional
 $\QYK_0$ takes the following form
\[ \varepsilon_{EM}(f)={\sqrt{-g}}
T{^0}{_0}(f)=\frac{\sqrt{-g}}{8\pi} (E^2+B^2)
 = \frac{\sqrt{-g^{00}}}{8\pi} \left( g^{kl}E_k E_l + \frac12
 B_{kl}B_{mn}g^{km}g^{ln} \right)  , \]
where the energy-momentum tensor $T$ is given by \eq{TEM},
$E_k=f_{0k}$ and $B_{kl}=f_{kl}$.
Formula \eq{TEM} implies that the tensor $f$ and
 its dual field
{\setlength\arraycolsep{2pt}
\begin{eqnarray}\label{Kerrfstar}
\ast f_{aux} : = F(R,Y)/4m & = &
\frac{a\sin\theta\left(r^2-a^2\cos^2\theta\right)} {\rho ^4} \rd
t\wedge \rd\theta + \frac{2ar\cos\theta} {\rho ^4}\rd t\wedge \rd
r \nonumber\\
& & {}+
\frac{\sin\theta\left(r^2+a^2\right)\left(r^2-a^2\cos^2\theta\right)}
{\rho ^4}\rd\theta \wedge \rd\phi + {}\nonumber\\
& & {} + \frac{2a^2r\cos\theta\sin^2\theta}{\rho ^4}\rd r \wedge
\rd \phi\,
\end{eqnarray}}
 give the same density $\varepsilon_{EM}$.

For $f_{aux}$ given by \eq{Kerrf} we get the following non-vanishing
components:
\[ E_r=\frac{r^2-a^2\cos^2\theta} {\rho ^4} \; , \quad
E_\theta = -\frac{2a^2r\sin\theta\cos\theta}{\rho ^4} \, ,\]
\[ B_{r\phi}= \frac{a\sin^2\theta(r^2-a^2\cos^2\theta)}{\rho ^4} \, , \quad
B_{\theta\phi}= -\frac{2ar\sin\theta\cos\theta(r^2+a^2)} {\rho^4} \, ,
\]
and finally
\be\label{density}  \varepsilon_{EM}(f)= \frac{\sin\theta}{8\pi\rho^2}
{(r^2+a^2)^2 +a^2 \Delta \sin^2\theta  \over
 (r^2+a^2)^2 -a^2 \Delta \sin^2\theta} \, .\ee
Our aim is to show that the ``spin energy density''
coming from functional \eq{EMV}
is just the rescaled density \eq{density}.
For this purpose
let us consider another Maxwell field
(proportional to $f$) \[
F_{rot}= C(\alpha)\frac{a}{\sqrt2}f_{aux} \, , \]
 where $\alpha:=a/m$ and
the rescaling function $C(\alpha)$ will be specified in the sequel.
Obviously \eq{density} gives
\be\label{spinen} \varepsilon_{rot}(F)= \frac{C^2a^2\sin\theta}{16\pi\rho^2}
{(r^2+a^2)^2 +a^2 \Delta \sin^2\theta  \over
 (r^2+a^2)^2 -a^2 \Delta \sin^2\theta} \, .\ee
Integrating density $\varepsilon_{rot}(F)$ outside of the horizon
we obtain a total ``spin energy'':
\be\label{Espin} M_{rot} = 2\pi
\int_{r_{+}}^{\infty}\!\rd r \int_{0}^{\pi}\!\rd\theta \, \varepsilon_{rot}
\, .
\ee
Using in the integral \eq{Espin} new variables $y=r/m$ and $x=\cos\theta$
it is easy to rewrite it in the form
\[ M_{rot} = \frac{C^2 a^2}{8m} \Psi(\alpha) \, ,\]
where
\be\label{Psint} \Psi(\alpha) :=
\int_{1+\sqrt{1-\alpha^2}}^{\infty}\rd y \int_{-1}^{1}
\frac{\rd x}{y^2+\alpha^2 x^2}
{(y^2+\alpha^2)^2 +\alpha^2 (y^2-2y+\alpha^2) (1-x^2)  \over
 (y^2+\alpha^2)^2 -\alpha^2 (y^2-2y+\alpha^2) (1-x^2)} \, .
\ee
The ``constant'' $C$ is defined by the relation
which expresses a balance of the energy:
\be\label{balen} m = M_{irr}+ M_{rot} \, .\ee
 Hence
\be\label{C2} C^2 = {8\left( 1-\sqrt{1+\sqrt{1-\alpha^2}\over 2} \right)\over
\alpha^2 \Psi }\, .\ee
The integral \eq{Psint}, which is not elementary, may be easily
approximated with the help of symbolic software like {\sc Mathematica} or
{\sc Waterloo Maple} and it gives as follows
\[ \Psi(\alpha) = 1 + \frac{\alpha^2}{4} + \frac{33\,\alpha^4}{280} +
\frac{3059\,\alpha^6}{43200} + \frac{835171\,\alpha^8}{17297280}
+ O(\alpha^9) \, , \]
\[ {8\left( 1-\sqrt{1+\sqrt{1-\alpha^2}\over 2} \right)\over \alpha^2 } =
   1 + \frac{5\,\alpha ^2}{16} + \frac{21\,\alpha ^4}{128} +
   \frac{429\,\alpha ^6}{4096} + \frac{2431\,\alpha ^8}{32768}
   + O(\alpha^{10})  \]
   and
\[ C = 1 + \frac{\alpha^2}{32} + \frac{1061\,\alpha^4}{71680} +
\frac{397907\,\alpha^6}{44236800} + \frac{245297388031\,\alpha^8}{39675808972800} +
  O(\alpha^9) \, . \]
However, for $\alpha=0$ the integral \eq{Psint} is elementary
and gives $\Psi(0)=1=C(0)$.\\
Formula \eq{spinen} together with \eq{C2} defines
uniquely the ``spin energy density'' coming from \eq{EMV} and
compatible with \eq{balen}.


\appendix
\section{Conformal rescaling of CYK tensors}\label{conf_resc_ch}
In this section we will be dealing with conformal transformations
and their impact on conformal Yano-Killing tensors. Since most of
the considerations here are independent on the dimension of a
manifold, we will not restrict ourselves to spacetime of dimension
four. We assume that we are dealing with $n$-dimensional manifold
and that this manifold has a metric $g$ (signature of $g$ plays no
role).

Let
$\Gamma^\alpha{}_{\mu\nu}$ denotes Christoffel symbols of Levi-Civita
connection associated with the metric $g$. We have:
\begin{equation}\label{conn_coef}
    \Gamma^\alpha{}_{\mu\nu} = \frac12 g^{\alpha\beta}(g_{\beta\mu,\nu}
    + g_{\beta\nu,\mu} - g_{\mu\nu,\beta}).
\end{equation}
Let $\tilde{g}$ be conformally rescaled metric, i.e.
$\tilde{g}_{\mu\nu} :=\Omega^2 g_{\mu\nu}$ (and what follows,
$\tilde{g}^{\mu\nu}:=\Omega^{-2} g^{\mu\nu}$), where $\Omega$ is a
certain positive function ($\Omega>0$). We will denote Christoffel
symbols of this metric by $\tilde{\Gamma}^\alpha{}_{\mu\nu}$,
and covariant derivative associated with them by
$\tilde{\nabla}_\mu$. Obviously, for $\tilde{g}_{\mu\nu}$ and
$\tilde{\Gamma}^\alpha{}_{\mu\nu}$ we have {\setlength\arraycolsep{2pt}
\begin{eqnarray}\label{conn_coef_conf}
    \tilde{\Gamma}^\alpha{}_{\mu\nu} & = & \frac12 \tilde{g}^{\alpha\beta}
    (\tilde{g}_{\beta\mu,\nu} + \tilde{g}_{\beta\nu,\mu} - \tilde{g}_{\mu\nu,\beta})
    \nonumber\\[6pt]
    & = & \frac12 \Omega^{-2}g^{\alpha\beta}\left((\Omega^2g_{\beta\mu}){}_{,\nu})
    + (\Omega^2g_{\beta\nu}){}_{,\mu} + (\Omega^2g_{\mu\nu}){}_{,\beta}\right)
    \nonumber\\[6pt]
    & = & \Gamma^\alpha{}_{\mu\nu} + \frac12\Omega^{-2}g^{\alpha\beta}
    \left((\Omega^2)_{,\nu}g_{\beta\mu} + (\Omega^2)_{,\mu}g_{\beta\nu}
    + (\Omega^2)_{,\beta}g_{\mu\nu}\right)\nonumber\\[6pt]
    & = & \Gamma^\alpha{}_{\mu\nu} + g^{\alpha\beta} \left(U_{,\nu}g_{\beta\mu}
    + U_{,\mu}g_{\beta\nu} + U_{,\beta}g_{\mu\nu}\right)\nonumber\\[8pt]
    & = & \Gamma^\alpha{}_{\mu\nu} + \delta^\alpha{}_\mu U_{,\nu} + \delta^\alpha{}_{\nu}
    U_{,\mu} - g^{\alpha\beta}U_{,\beta}g_{\mu\nu}\,,
\end{eqnarray} }
which is analogous
to~(\ref{conn_coef}). Here $U:=\log\Omega$.
 Using the formula~(\ref{conn_coef_conf})
and the following identities:
\begin{equation}\label{cov_diff_def}
    \nabla_\mu X_{\nu\rho} = X_{\nu\rho,\mu} - X_{\alpha\rho}
    \Gamma^\alpha{}_{\nu\mu} - X_{\nu\alpha}\Gamma^\alpha{}_{\rho\mu}\, ,
\end{equation}
\begin{equation}\label{cov_diff_def_conf}
    \tilde{\nabla}_\mu X_{\nu\rho} = X_{\nu\rho,\mu} - X_{\alpha\rho}
    \tilde{\Gamma}^\alpha{}_{\nu\mu} - X_{\nu\alpha}\tilde{\Gamma}^\alpha{}_{\rho\mu}
\end{equation}
(which are true for any tensor $X_{\mu\nu}$), we get:
{\setlength\arraycolsep{2pt}
\begin{eqnarray}\label{cov_der_conf_resc}
    \tilde{\nabla}_\mu X_{\nu\rho} & = & \nabla_\mu X_{\nu\rho} - X_{\mu\rho}U_{,\nu}
    - X_{\nu\mu}U_{,\rho} - 2X_{\nu\rho}U_{,\mu} + {}\nonumber\\[8pt]
    & & {}  + g^{\alpha\beta}U_{,\beta}\left( X_{\alpha\rho}g_{\mu\nu}
    + X_{\nu\alpha}g_{\mu\rho}\right).
\end{eqnarray}}

We will now prove the formula~(\ref{conf_resc}). Let us introduce
the following notation:
\begin{equation}\label{Q_conf}
    \tilde{Q}_{\mu\nu} = \Omega^3 Q_{\mu\nu}\,,
\end{equation}
\begin{equation}\label{div_Q_conf}
    \tilde{\xi}_{\rho}:=\tilde{g}^{\mu\nu}\tilde{\nabla}_{\mu}\tilde{Q}_{\nu\rho}\,.
\end{equation}
Using the formula~(\ref{cov_der_conf_resc}), we can rewrite the
formula~(\ref{div_Q_conf}) in the following form:
{\setlength\arraycolsep{2pt}
\begin{eqnarray}\label{div_Q_conf_scal}
    \tilde{\xi}_\rho & = & \tilde{g}^{\mu\nu}\nabla_{\mu}\tilde{Q}_{\nu\rho}
    - 3\tilde{g}^{\mu\nu} \tilde{Q}_{\mu\rho}U_{,\nu} + \Omega^{-2}g^{\mu\nu}g^{\alpha\beta}
    U_{,\beta}\left( \tilde{Q}_{\alpha\rho}g_{\mu\nu} + \tilde{Q}_{\nu\alpha}
    g_{\mu\rho}\right)\nonumber\\
    & = &\tilde{g}^{\mu\nu}\nabla_{\mu}\tilde{Q}_{\nu\rho} - 3\tilde{g}^{\mu\nu}
    \tilde{Q}_{\mu\rho}U_{,\nu} + \tilde{g}^{\alpha\beta} U_{,\beta}
    \left( n\tilde{Q}_{\alpha\rho} + \tilde{Q}_{\nu\alpha} \delta^\nu{}_{\rho}\right)
    \nonumber\\
    & = & \tilde{g}^{\mu\nu}\nabla_{\mu}\tilde{Q}_{\nu\rho} + (n-4)\tilde{g}^{\mu\nu}
    \tilde{Q}_{\mu\rho}U_{,\nu}\nonumber\\
    & = & \Omega^{-2}g^{\mu\nu}\nabla_{\mu}(\Omega^3Q_{\nu\rho}) + (n-4)\Omega
    g^{\mu\nu} Q_{\mu\rho}U_{,\nu}\nonumber\\
    & = & \Omega g^{\mu\nu}\nabla_{\mu}Q_{\nu\rho} + \Omega^{-2}g^{\mu\nu}(\Omega^3)_{,\mu}
    Q_{\nu\rho} + (n-4)\Omega g^{\mu\nu} Q_{\mu\rho}U_{,\nu}\nonumber\\
    & = & \Omega\left( \xi_\rho + (n-1)g^{\mu\nu} Q_{\mu\rho}U_{,\nu}\right),
\end{eqnarray}}where $\xi_\rho:=g^{\mu\nu}\nabla_{\mu}Q_{\nu\rho}$.
The last equality in~(\ref{div_Q_conf_scal}) is a result of the following identities:
$\Omega^{-2}(\Omega^3)_{,\mu}=3\Omega_{,\mu}=3\Omega
(\log\Omega)_{,\mu}=3\Omega U_{,\mu}$. Using the
formula~(\ref{cov_der_conf_resc}) for~$\tilde{Q}$ we get:
{\setlength\arraycolsep{2pt}
\begin{eqnarray}\label{conf_resc2}
    \tilde{\nabla}_\mu\tilde{Q}_{\nu\rho} & = & \Omega^3\Big[\nabla_\mu Q_{\nu\rho}
    + Q_{\nu\rho}U_{,\mu} - Q_{\nu\mu}U_{,\rho} - Q_{\mu\rho}U_{,\nu}\nonumber\\
    & & + g^{\alpha\beta}U_{,\beta}(g_{\mu\nu}Q_{\alpha\rho}+g_{\mu\rho}Q_{\nu\alpha})
    \Big]
\end{eqnarray}}
and what follows: {\setlength\arraycolsep{2pt}
\begin{eqnarray}\label{conf_resc3}
    \tilde{\nabla}_\mu\tilde{Q}_{\nu\rho} + \tilde{\nabla}_\rho\tilde{Q}_{\nu\mu}
    & = & \Omega^3\Big[\nabla_\mu Q_{\nu\rho} + \nabla_\rho Q_{\nu\mu}
    + g^{\alpha\beta}U_{,\beta}(g_{\mu\nu}Q_{\alpha\rho} + {}\nonumber\\
    & & {} + g_{\rho\nu}Q_{\alpha\mu}
    +2g_{\mu\rho}Q_{\nu\alpha})\Big] \, .
\end{eqnarray}}
Formulae (\ref{CYK_eq1}),
 (\ref{div_Q_conf_scal}) and (\ref{conf_resc3})
imply that
{\setlength\arraycolsep{2pt}
\begin{eqnarray}\label{conf_resc_proof}
    {\cal Q}_{\mu\nu\rho}(\tilde{Q}, \tilde{g}) & = & \tilde{\nabla}_\rho\tilde{Q}_{\mu\nu}
    + \tilde{\nabla}_\mu\tilde{Q}_{\rho\nu} - \frac2{n-1}\left(\tilde{g}_{\mu\rho}
    \tilde{\xi}_\nu-\tilde{g}_{\nu(\mu}\tilde{\xi}_{\rho)}\right)\nonumber\\
    & = & \Omega^3 \left[\nabla_\rho Q_{\mu\nu} + \nabla_\mu Q_{\rho\nu}
    - \frac2{n-1}\left(g_{\mu\rho}\xi_\nu-g_{\nu(\mu}\xi_{\rho)}\right)\right]\nonumber\\
    & = & \Omega^3 {\cal Q}_{\mu\nu\rho}(Q, g),
\end{eqnarray}}
which proves the formula~(\ref{conf_resc}).

\section{Integrability conditions for the CYK equation}\label{war_calk_dod}
For a manifold $M$ equipped with metric tensor $g_{\mu\nu}$ and for
any tensor field $Q_{\mu\nu}$ on $M$ the following equality holds
\begin{equation}\label{Riemann_id}
    Q_{\lambda\kappa;\nu\mu} - Q_{\lambda\kappa;\mu\nu} = Q_{\sigma\kappa}
    R^{\sigma}{}_{\lambda\nu\mu}
    + Q_{\lambda\sigma}R^{\sigma}{}_{\kappa\nu\mu}\ ,
\end{equation}
where $R^{\mu}{}_{\nu\rho\sigma}$ is Riemann tensor of metric
$g_{\mu\nu}$. Obviously we assume that $Q$ is skew-symmetric
which implies that
\begin{equation}\label{divxi}
2\xi^{\mu}{}_{;\mu}=Q^{\lambda\kappa}{}_{;\lambda\kappa} - Q^{\lambda\kappa}{}_{;\kappa\lambda}
= 2Q^{\kappa\sigma}R_{\sigma\kappa} = 0.
\end{equation}

Changing the names of indices we write \eq{Riemann_id} three
times:
\[
Q_{\lambda\kappa;\nu\mu} - Q_{\lambda\kappa;\mu\nu} =
Q_{\sigma\kappa} R^{\sigma}{}_{\lambda\nu\mu} +
Q_{\lambda\sigma}R^{\sigma}{}_{\kappa\nu\mu}\ ,
\]
\[
Q_{\mu\kappa;\lambda\nu} - Q_{\mu\kappa;\nu\lambda} =
Q_{\sigma\kappa} R^{\sigma}{}_{\mu\lambda\nu} +
Q_{\mu\sigma}R^{\sigma}{}_{\kappa\lambda\nu}\ ,
\]
\[
Q_{\nu\kappa;\mu\lambda} - Q_{\nu\kappa;\lambda\mu} =
Q_{\sigma\kappa} R^{\sigma}{}_{\nu\mu\lambda} +
Q_{\nu\sigma}R^{\sigma}{}_{\kappa\mu\lambda}\ .
\]
We add the first equation, subtract the second one and finally add the third
equation. Hence we obtain
\begin{eqnarray} \nonumber
& & \hspace*{-1cm} Q_{\lambda\kappa;\nu\mu} - Q_{\lambda\kappa;\mu\nu} -
Q_{\mu\kappa;\lambda\nu} + Q_{\mu\kappa;\nu\lambda} +
Q_{\nu\kappa;\mu\lambda} - Q_{\nu\kappa;\lambda\mu}
 \\[6pt] \nonumber
& = & 2Q_{\lambda\kappa;\nu\mu} - (Q_{\lambda\kappa;\mu} +
Q_{\mu\kappa;\lambda})_{;\nu} + (Q_{\mu\kappa;\nu} +
Q_{\nu\kappa;\mu})_{;\lambda} - (Q_{\nu\kappa;\lambda}
 + Q_{\lambda\kappa;\nu})_{;\mu} \\[6pt]
& = & Q_{\sigma\lambda}R^{\sigma}{}_{\kappa\mu\nu} +
Q_{\sigma\mu}R^{\sigma}{}_{\kappa\lambda\nu} +
Q_{\sigma\nu}R^{\sigma}{}_{\kappa\lambda\mu} +
2Q_{\sigma\kappa}R^{\sigma}{}_{\mu\nu\lambda} \, .
\end{eqnarray}
In the last equality we used the standard property of curvature tensor:
$R^{\sigma}{}_{[\mu\nu\lambda]}=0$. Definition~(\ref{CYK_eq1})
applied to the terms in brackets in the above formula
implies
 {\setlength\arraycolsep{2pt}
\begin{eqnarray}\label{war_calk}
    2Q_{\lambda\kappa;\nu\mu} & = & \frac{2}{n-1}\left(g_{\lambda\mu}\xi_{\kappa;\nu}
    + g_{\nu\lambda}\xi_{\kappa;\mu} - g_{\mu\nu}\xi_{\kappa;\lambda}
    - g_{\kappa(\lambda}\xi_{\mu);\nu}
    + g_{\kappa(\mu}\xi_{\nu);\lambda} - g_{\kappa(\nu}\xi_{\lambda);\mu}\right)
    \nonumber\\
    & &  + Q_{\sigma\lambda}R^{\sigma}{}_{\kappa\mu\nu}
    + Q_{\sigma\mu}R^{\sigma}{}_{\kappa\lambda\nu}
     + Q_{\sigma\nu}R^{\sigma}{}_{\kappa\lambda\mu}
    + 2Q_{\sigma\kappa}R^{\sigma}{}_{\mu\nu\lambda}
    \nonumber\\
    & & {} + {\cal Q}_{\lambda\kappa\mu; \nu}
    - {\cal Q}_{\mu\kappa\nu ; \lambda}
    + {\cal Q}_{\nu\kappa\lambda ; \mu} \, ,
\end{eqnarray}}
where the covector $\xi_\mu = \nabla^\lambda Q_{\lambda\mu}$ (cf. \eq{div_Q}).
 Hence for a CYK tensor $Q$ we obtain \eq{war_calk1}.

Contracting \eq{war_calk} with respect to indices $\mu$ and $\nu$
and using the basic properties \eq{wlQ} of $\cal Q$ we get
\be\label{warcalkzQ}
Q_{\lambda\kappa}{^{;\mu}}{_\mu}
+ R^{\sigma}{}_{\kappa\lambda\mu}Q{^\mu}{_{\sigma}}+ Q_{\sigma\kappa}R^\sigma{_\lambda}
 +\frac{2}{n-1}\xi_{(\kappa;\lambda)} + \frac{1}{n-1}g_{\kappa\lambda}
 \xi^\mu{_{;\mu}} =
 \nabla^\mu{\cal Q}_{\mu\kappa\lambda} -\frac{n-4}{n-1}\xi_{\kappa;\lambda} \, .
\ee Formulae~\eq{almost_killing} and \eq{warcalkzQ} imply
 the following equation for a CYK tensor $Q$:
\be\label{warcalkbis}
\nabla^\mu\nabla_\mu Q_{\lambda\kappa} =
R^{\sigma}{}_{\kappa\lambda\mu}Q_{\sigma}{^\mu}
 -R_{\sigma[\kappa}Q_{\lambda]}{^\sigma} \,
\ee
for $\dim M=4$.
It is interesting to point out that for compact four-dimensional
Riemannian manifolds
equation~\eq{warcalkbis} is equivalent to Definition \ref{CYK_df}:
\begin{Theorem}\label{th9}
Let $M$ be a compact (without boundary) Riemannian manifold with
$\dim M = 4$.
Then a two-form $Q$
is a CYK tensor iff \ 
 \[ \nabla^\mu\nabla_\mu Q_{\lambda\kappa} =
R^{\sigma}{}_{\kappa\lambda\mu}Q_{\sigma}{^\mu} 
 + R_{\sigma[\lambda}Q_{\kappa]}{^\sigma}\, . \]
\end{Theorem}
\begin{proof}
We need to show that equation~\eq{warcalkbis} implies
${\cal Q}_{\lambda\kappa\mu}(Q,g)=0$.

Similarly to~\eq{almost_killing} we derive
\[ \frac{2}{3}\xi_{(\mu;\lambda)}+\frac{1}{3}g_{\mu\lambda}
 \xi^\nu{_{;\nu}} -
    R_{\sigma(\mu} Q_{\lambda)}{}^{\sigma}
    +\frac12\nabla^\sigma{\cal Q}_{\lambda\sigma\mu}
    =0  \, , \quad  4\xi^{\mu}{_{;\mu}}
    +\nabla^\sigma{\cal Q}_{\nu\sigma\mu}g^{\mu\nu}
    =0 \, , \]
which together with \eq{divxi} and \eq{warcalkzQ} gives\footnote{For a CYK tensor $Q$
 formula \eq{wcRQ}  obviously gives equation~\eq{warcalkbis}.}
\be\label{wcRQ}
\nabla^\mu\nabla_\mu Q_{\lambda\kappa} +
R^{\sigma}{}_{\kappa\lambda\mu}Q{^\mu}{_{\sigma}} 
+R_{\sigma[\kappa}Q_{\lambda]}{^\sigma} =
 \nabla^\mu{\cal Q}_{\mu\kappa\lambda} 
  + \frac12 \nabla^\sigma{\cal Q}_{\kappa\sigma\lambda} 
  \, .
\ee
Contracting the above equality with $Q$ and
assuming equation~\eq{warcalkbis} we get
\begin{eqnarray}\nonumber 0 = \left(\nabla^\mu{\cal Q}_{\mu\kappa\lambda}
  + \frac12\nabla^\sigma{\cal Q}_{\lambda\sigma\kappa}\right)Q^{\kappa\lambda}
  & = & \nabla^\mu \left({\cal Q}_{\mu\kappa\lambda}Q^{\kappa\lambda}\right)-
  {\cal Q}_{\mu\kappa\lambda}\nabla^\mu Q^{\kappa\lambda} \\ &= &
 \nabla^\mu \left({\cal Q}_{\mu\kappa\lambda}Q^{\kappa\lambda}\right)
 + \frac12 {\cal Q}_{\lambda\kappa\mu} {\cal Q}^{\lambda\kappa\mu} \, .
  \end{eqnarray}
  Finally, we integrate the above formula over $M$,
  a total divergence drops out, and the integral
  $\displaystyle \int_M \sqrt{\det g}{\cal Q}_{\lambda\kappa\mu}
  {\cal Q}^{\lambda\kappa\mu}$ vanishes.
  This implies ${\cal Q}^{\lambda\kappa\mu}=0$.
\end{proof}
A similar result holds for a $p$-form $Q$ in $2p$-dimensional $M$
(see \cite{Semmelmann}). 

\subsection{Miscellaneous results}

It is worth pointing out that equation \eq{warcalkbis}
can be derived\footnote{The calculations are similar to
the proof of Theorem \ref{th9}.}
from a variational principle
$\displaystyle \delta \int_M {\cal L} =0$,
where $\displaystyle {\cal L}[Q]=\frac14\sqrt{\det g}
{\cal Q}_{\lambda\kappa\mu} {\cal Q}^{\lambda\kappa\mu}(Q,g)$.

Equation \eq{warcalkbis} for Ricci flat metrics gives
\[ 2\nabla^\mu\nabla_\mu Q_{\kappa\lambda} =
 2R^{\sigma}{_{\kappa\lambda\mu}}Q{^\mu}{_{\sigma}}=
 R_{\kappa\lambda\sigma\mu}Q^{\sigma\mu} = F_{\kappa\lambda} (W,Q) \]
and, in particular, it is true for $Q=Y$ in Kerr spacetime.
Moreover,
\[ \frac 14 F_{\mu\nu}(W,Y) Y^{\mu\nu} =
\frac{2mr}{\rho^2} = \frac{2mr}{r^2+a^2\cos^2\theta}=g(\partial_t,\partial_t)+1
\]
 gives a nice density in Boyer-Lindquist coordinates:
 \[ \sqrt{-g}F_{\mu\nu}(W,Y) Y^{\mu\nu}= 8mr\sin\theta \, . \]

A complex-valued function $\zeta:= r+\imath \cos\theta$
enables one to write some objects in Kerr spacetime in a compact form
like the Ernst potential ${\cal E} = 1 - {2m}/{\bar\zeta}$
or Newman--Penrose scalar $\Psi_2= - {m}/{\bar\zeta^3}$ (cf. \cite{SKM}),
where by $\bar\zeta$ we denote the complex conjugate i.e.
$\bar\zeta:= r-\imath \cos\theta$. In this context
in Eddington--Finkelstein coordinates we have
\[ Y+\imath \ast Y = -\imath\zeta\rd\zeta\wedge(\rd v - a\sin^2\theta\rd
\tilde{\phi}) + \bar\zeta\zeta^2\sin\theta\rd\theta\wedge\rd\tilde{\phi} \, ,\]
\begin{eqnarray}
K'+\imath K'' & = & \frac{\bar\zeta -\zeta}{2\zeta} \left(
a^2\sin^2\theta \, \partial_v\!\otimes\partial_v +
\partial_\theta\!\otimes\partial_\theta +
\frac{1}{\sin^2\theta}
\partial_{\scriptscriptstyle\tilde{\phi}}\!\otimes
\partial_{\scriptscriptstyle\tilde{\phi}} + 2a \partial_v \!\otimes_{s}\!
\partial_{\scriptscriptstyle\tilde{\phi}} \right)
\nonumber \\[10pt]
& & + \frac{\bar\zeta +\zeta}{2\zeta} \left( \Delta
\partial_r\!\otimes\partial_r +
2(r^2+a^2)\partial_v \!\otimes_{s}\! \partial_r
+ 2a \partial_r \!\otimes_{s}\!
\partial_{\scriptscriptstyle\tilde{\phi}} \right)\, , \nonumber
\end{eqnarray}
and, in particular, $\tr \left(K' +\imath K''\right) = 2 {\bar\zeta}^2=
\frac{{\cal E} - 1}{\Psi_2}$.

\end{document}